%% file: EASE18.tex
\documentclass[sigconf]{acmart}
\usepackage{fancybox}
\usepackage[export]{adjustbox}
\newcommand\mybox[2][]{\tikz[overlay]\node[fill=gray,inner sep=1pt, anchor=text, rectangle, rounded corners=0.5mm,#1] {#2};\phantom{#2}}
\usepackage{pifont} %For Dings :-)
\usepackage{subfig}
\usepackage{colortbl}
\usepackage{hhline} 
\usepackage{enumitem}
\usepackage{tikz}

\usetikzlibrary{patterns}
\tikzset{ 
table/.style={
  matrix of nodes,
  row sep=-\pgflinewidth,
  column sep=-\pgflinewidth,
  nodes={draw,rectangle,text width=2cm,align=center},
  text depth=1.5ex,
  text height=4.5ex,
  nodes in empty cells
}
}

\tikzset{
    every picture/.style={
        remember picture,   % Make nodes available to all TikZ pictures
        inner xsep=0pt, % Remove horizontal padding
        inner ysep=1pt, % Set small vertical padding
        baseline,       % Align TikZ pictures at the baseline
        every node/.style={
            anchor=base % Align all nodes at the baseline
        }
    }
}

\usepackage{booktabs} % For formal tables

\usepackage{listings}
\usepackage{color}

\definecolor{dkgreen}{rgb}{0,0.5,0}
\definecolor{gray}{RGB}{224,224,224}
\definecolor{mauve}{rgb}{0.58,0,0.82}
\definecolor{lb}{RGB}{204,255,229}

\usepackage{framed}
\usepackage{graphicx,multirow}

\setcopyright{rightsretained}
\usepackage{booktabs} % For formal tables
\newcommand{\ignore}[1]{}
\usepackage{xcolor,colortbl}
\definecolor{G1}{RGB}{249, 246, 246}
\definecolor{light}{RGB}{189,213,215}
\definecolor{blue1}{RGB}{153,205,255}
\definecolor{g1}{RGB}{195, 207, 222}
\definecolor{blue2}{RGB}{0,128,255}
\definecolor{blue3}{RGB}{0,76,153}
\definecolor{red1}{RGB}{255,180,180}
\definecolor{red2}{RGB}{255,95,95}
\definecolor{red3}{RGB}{205,0,0}
\definecolor{green}{RGB}{229,255,204}
\definecolor{red}{rgb}{0.85,0.8,0.9}			
\usepackage{etoolbox}
\makeatletter
\patchcmd{\maketitle}{\@copyrightspace}{}{}{}
\makeatother
\setcopyright{rightsretained}
\usepackage{booktabs}
\usepackage{courier}
\usepackage{fixltx2e}

\usepackage{mdframed}
\setcopyright{none}
\setlist[itemize]{leftmargin=7mm}

\begin{document}

\title{Task Interruption in Software Development Projects}
\subtitle{What Makes some Interruptions More Disruptive than Others?}
%\subtitlenote{The full version of the author's guide is available as
 % \texttt{acmart.pdf} document}

\author{Zahra Shakeri Hossein Abad$^{\textasteriskcentered}$, Oliver Karras$^{\dagger}$, Kurt Schneider$^{\dagger}$, Ken Barker$^{\textasteriskcentered}$, Mike Bauer$^{\ddagger}$}
%\authornote{--}
\affiliation{%
\institution{$^{\textasteriskcentered}$ Department of Computer Science, University of Calgary, Canada, \{zshakeri, kbarker\}@ucalgary.ca}
 \institution{{${\dagger}$ {\normalsize Software Engineering Group, Leibniz Universit\"at Hannover, Germany, \{oliver.karras, kurt.schneider\}@inf.uni-hannover.den}}}
   \institution{${\ddagger}$ Arcurve Inc., Calgary, Canada, mike.bauer@arcurve.com}
%  \streetaddress{}
%  \city{Calgary} 
%  \state{AB} 
}

%
%\author{Zahra Shakeri Hossein Abad, Ken Barker}
%%\authornote{--}
%\orcid{1234-5678-9012}
%\affiliation{%
%  \institution{Department of Computer Science \\ \small University of Calgary, Canada}
%%  \streetaddress{}
%%  \city{Calgary} 
%%  \state{AB} 
%}
%\email{zshakeri@ucalgary.ca}
%%
%%
%\author{Oliver Karras, Kurt Schneider}
%\affiliation{\institution{Software Engineering Group\\ \small Leibniz 
%Universit\"at Hannover, Germany}}
%\email{{ oliver.karras, kurt.schneider@inf.uni-hannover.de}}
%
%\author{Mike Bauer}
%\affiliation{\institution{Arcurve Inc., \\ \small Calgary, Canada}}
%\email{bauer@arcurve.ca}

%\author{Mike Bauer}
%\affiliation{\institution{Arcurve Inc., Calgary, Canada}}
%\email{bauer@arcurve.cat}
%% The default list of authors is too long for headers}
%\renewcommand{\shortauthors}{S.H Abad et al.}

\begin{abstract}
Multitasking has always been an inherent part of software development and is known as the primary source of interruptions due to task switching in software development teams. Developing software involves a mix of analytical and creative work, and requires a significant load on brain functions, such as working memory and decision making. Thus, task switching in the context of software development imposes a cognitive load that causes software developers to lose focus and concentration while working thereby taking a toll on productivity. To investigate the disruptiveness of task switching and interruptions in software development projects, and to understand the reasons for and perceptions of the disruptiveness of task switching we used a mixed-methods approach including a longitudinal data analysis on 4,910 recorded tasks of 17 professional software developers, and a survey of 132 software developers. We found that, compared to task-specific factors (e.g. priority, level, and temporal stage), contextual factors such as interruption type (e.g. self/external), time of day, and task type and context are a more potent determinant of task switching disruptiveness in software development tasks. Furthermore, while most survey respondents believe external interruptions are more disruptive than self-interruptions, the results of our retrospective analysis reveals otherwise. We found that self-interruptions (i.e. voluntary task switchings) are more disruptive than external interruptions and have a negative effect on the performance of the interrupted tasks. Finally, we use the results of both studies to provide a set of comparative vulnerability and interaction patterns which can be used as a mean to guide decision-making and forecasting the consequences of task switching in software development teams.

%Our work brings out a connection between researchers and practitioners
\end{abstract}

%
% The code below should be generated by the tool at
% http://dl.acm.org/ccs.cfm
% Please copy and paste the code instead of the example below. 
%
%\begin{CCSXML}
%<ccs2012>
% <concept>
%  <concept_id>10010520.10010553.10010562</concept_id>
%  <concept_desc>Computer systems organization~Embedded systems</concept_desc>
%  <concept_significance>500</concept_significance>
% </concept>
% <concept>
%  <concept_id>10010520.10010575.10010755</concept_id>
%  <concept_desc>Computer systems organization~Redundancy</concept_desc>
%  <concept_significance>300</concept_significance>
% </concept>
% <concept>
%  <concept_id>10010520.10010553.10010554</concept_id>
%  <concept_desc>Computer systems organization~Robotics</concept_desc>
%  <concept_significance>100</concept_significance>
% </concept>
% <concept>
%  <concept_id>10003033.10003083.10003095</concept_id>
%  <concept_desc>Networks~Network reliability</concept_desc>
%  <concept_significance>100</concept_significance>
% </concept>
%</ccs2012>  
%\end{CCSXML}

%\ccsdesc[500]{Computer systems organization~Embedded systems}
%\ccsdesc[300]{Computer systems organization~Redundancy}
%\ccsdesc{Computer systems organization~Robotics}
%\ccsdesc[100]{Networks~Network reliability}
%\settopmatter{printacmref=false} %To remove the ACM Author Format!

\acmYear{2018}
\acmConference[EASE'18-22nd International Conference on Evaluation and
Assessment in Software Engineering 2018]{}{June 28--29, 2018}{Christchurch,
New Zealand}
\acmPrice{}
\acmDOI{TBA}
\acmISBN{TBA}
\maketitle
\input{EASE-Body}

\bibliographystyle{ACM-Reference-Format}
\bibliography{sigproc} 

\end{document}

%% file: EASE-Body.tex
\section{Introduction}
%\Ko2007 supports need for more info

Software development has undergone significant changes over the past decade. Traditionally siloed development teams are more collaborative and included more stakeholders from more disciplines than ever before. The need for faster-time-to-market, frequent releases, continuous integration, and continuous delivery has made frequent task switching an unavoidable part of software development projects. \emph{Task switching}, commonly referred to as \emph{multitasking}~\cite{Sky} and \emph{interruption}~\cite{Mind} is the act of starting one task and moving to another before finishing the first. Developers often have to switch tasks for various reasons: getting sidetracked to other tasks;  getting stuck or bored by complex or lengthy repetitive tasks; receiving priority change requests from the management team; or even something as simple as a question from a co-worker.
In a recent study of interruptions Parnin and Rugaber~\cite{Parnin} analyzed development logs of 10,000 programming sessions from 86 programmers and found that in a typical day, a developer's work is fragmented into many short sessions (i.e 15-30 minutes), and a programmer often spends a significant amount of time (i.e. 15-30 minutes) reconstructing working context before resuming interrupted tasks.
To gain a better grasp of the behaviour of task switching in software development projects we conducted an investigation of 44,515 tasks (recorded between 2013 and 2017) of 23 professional software developers at SET GmbH~\footnote{https://www.set.de}, a leading provider of standard software for output management ~\footnote{This analysis has been conducted to justify our research goals and it is different from the main longitudinal study of this paper}. 
We found that developers switch about two-thirds (59\%) of their daily tasks from which 40\% require context switching, and they never resume 29\% of their interrupted/switched tasks. 
While task switching in some cases help developers be more productive, it imposes a cognitive load on them: frequent task switching typically results in severe performance costs by increasing response latencies and error rates~\cite{SMT, Sky}, and can cause an initial decrease in how quickly people perform post-switching tasks~\cite{Design2}.

Research into developers' productivity and multitasking provide evidence on how multitasking and interruptions can impact productivity in software development teams~\cite{Sky, Pro1, Life}.
However, very little work~\cite{Parnin, Parnin2} has investigated the factors that can make task switching more disruptive for different types of software development tasks (e.g. programming, test, architecture, UI, and deployment). 
Given the paucity of empirical studies on the disruptiveness of
task switching and interruption in software development projects, it remains unclear what factors make which types of task interruptions more disruptive than others. 
This paper reports on a mixed-methods study exploring and analyzing factors influencing the vulnerability of various types of software development tasks to interruptions. A multivariate longitudinal analysis was conducted to investigate disruptive factors, such as the interruption type (i.e. self/external), context switching, and interruption timing (i.e. daytime, task stage), and to then perform comparative and cross-factor analysis on the vulnerability of various software development tasks based on these factors.
Further, a survey of 132 professional software developers from different organizations (e.g. Microsoft, Tableau Software, Ericsson, Bosch, and Cisco) explored practitioners' perceptions of and reasons for task switching and the disruptiveness of these interruptions. 
These studies show that context switching (e.g. task type and project), the abstraction level (i.e. main task, sub-task) and the temporal stage (i.e. early, late) of the interrupted task, and the interruption type (i.e. self, external) significantly impact the disruptiveness of interruptions and task switching in software development tasks. 
In summary, this paper makes the following contributions:
\begin{itemize}
\item It models interruption characteristics and presents a longitudinal analysis of 4,910 task logs of 17 professional software developers to study the vulnerability of various development tasks to interruptions and to explore the disruptive impact of interruption characteristics on different tasks' types.
\item It presents a survey of 132 professional software developers to identify their perceptions of the concept and impact of task switching and interruptions in software development projects.
\item It provides a set of comparative disruptiveness as well as cross-factor interaction patterns that can be used to guide task switching and to predict and manage the cognitive load associated with various interruptions.
\end{itemize}

%=================================================================================
 
\section{Background}
This section first describes concepts related to task switching and interruption. 
We formulate the dependent and independent variables of this study and conclude this section by reviewing the related work on interruption analysis in software engineering.                                                        
%=================================================================================

\subsection{Terms and Concepts}
\label{sec:TC}
\begin{figure}
\includegraphics[scale=0.75]{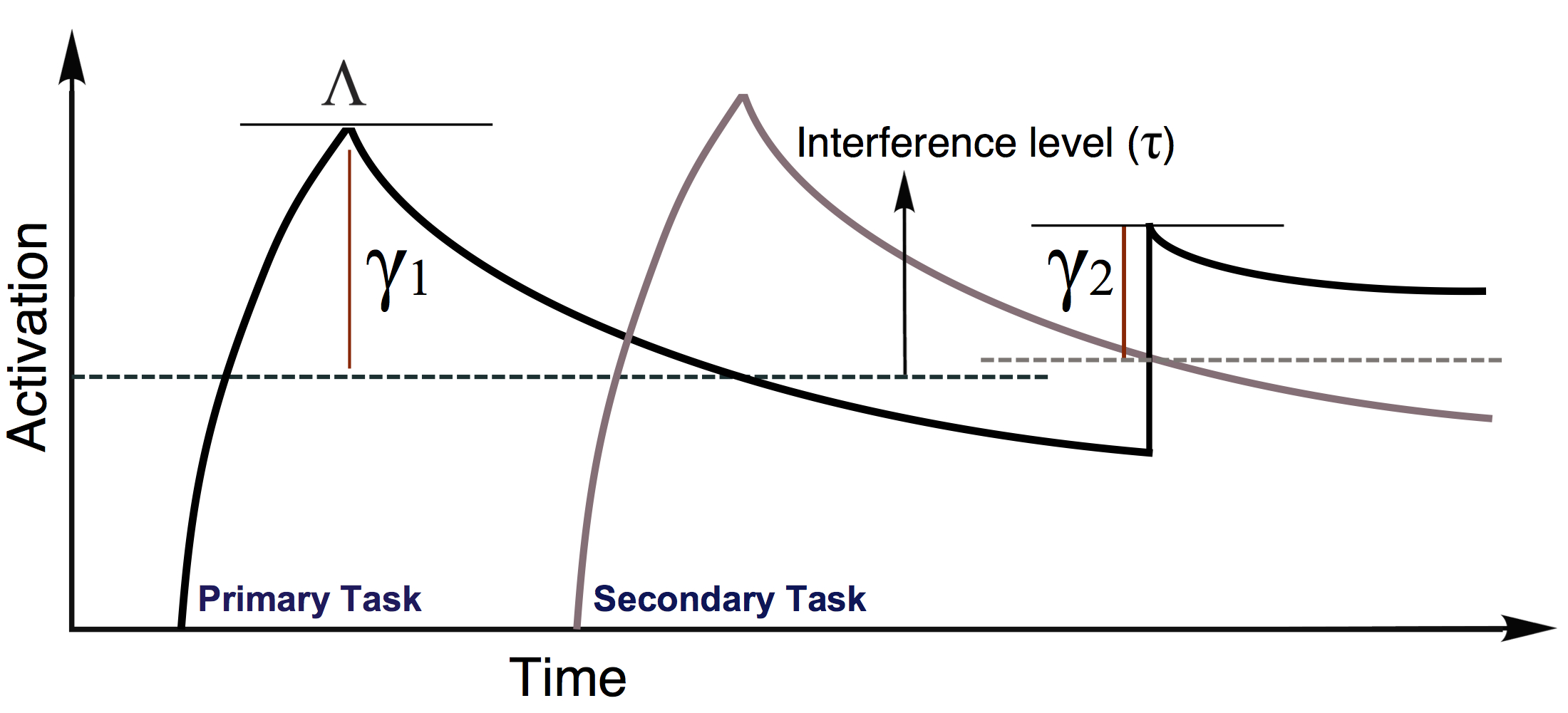}
\vspace{-3.75mm}
\caption{Goal activations in nested interruptions [adapted from Memory of Goals theory]. When \(\gamma=0\) the probability of recall is \%50.}
\label{fig:Fragments}
\vspace{-5mm}
\end{figure}

The information required to accomplish a task decays gradually in human memory, which results in a mental clutter of goals/tasks.
% This can have subsequent consequences for long-term memory encoding and goal retrieval \cite{MofGoals}.
  \mybox[fill=gray]{Problem state}, the main source of interference in multitasking environments, keeps track of task-related information that is not readily available in the external environment~\cite{Mind} or in the information associated with performing a task. 
Some tasks are reactive (e.g. answering an email or phone calls) and do not need to maintain a problem state.
Some tasks may utilize the problem state resource but do not need to maintain the information therein (e.g. stand-up meetings).
As interference only arises when the problem state resource is needed by two or more tasks, tasks that do not require problem state information will not experience interference on the problem state resource. 
Thus, we do not consider switching from (or interrupting) reactive tasks and tasks that do not need to maintain their problem state as \emph{task interruption}.
Instead, we refer to this type of task switching as \mybox[fill=gray]{no-task} interruptions. \mybox[fill=gray]{Activation} (\(\Lambda\)) or the momentary availability of the memory content controls the speed and reliability of access to the memory content after resuming a task \cite{Anderson}. 
As activation grows the information can be retrieved in a shorter amount of time \cite{Mind}. 
The time course of tasks activation in a sequential multitasking set-up is illustrated in Figure \ref{fig:Fragments}. The abscissa represents the time and the ordinate represents the activation level. 
The dashed line represents the \mybox[fill=gray]{Interference level} (\(\tau\)) (or activation threshold) and refers to the expected (mean) activation of the \emph{most} interrupting task~\cite{MofGoals}. 
\mybox[fill=gray]{Activation distance} (\(\gamma\)) represents the accuracy of memory for the current task and refers to the amount by which the resumed task at its peak is more active than the interference level. 
The memory-of-goals theory ~\cite{MofGoals} shows that the interference level depends on the number of interrupting tasks (nested interruptions) and the long-term durability (i.e. strength) of the information associated with these tasks. 
The more they are, or the stronger they are, the more they interfere with the target~\cite{MofGoals}, which contributes to a decrease in memory accuracy. 
For example, as illustrated in Figure~\ref{fig:Fragments}, the memory accuracy decreases as the number of interrupting tasks increases (\(\gamma_2<\gamma_1\)). 
The ACT-R theory computes the activation as a function of frequency of use (i.e. 
$\Lambda= \ln\big(\frac{n}{\sqrt[]{T}}\big)$), where \(n\) is the total number of times the memory item has been retrieved in its lifetime, and \(T\) is the length of this lifetime. 
%Activation plays a crucial role in probability and latency of recall after resuming an interrupted task. 
ACT-R formulates the probability of recall as an exponential function of activation distance (i.e. \(P_{recall} = \frac{1}{1+e^{-\gamma/s}}\)). 
Thus, as time passes without using an item, \(T\) for that item grows, whereas $n$ does not, producing decay (a decrease in the activation level). Given that activation (\(\Lambda\)) decreases by time and the activation threshold (\(\tau\)) (or the interference level) increases by the length of the interruptions and the number of distractors, we can conclude that the probability of recall decreases as a power function of \emph{time} and \emph{the number of distractors} \cite{Atomic}. Thus, in this paper, we study the vulnerability of software development tasks by exploring the impact of various interruption characteristics on these two dependent variables: (1) \mybox[fill=gray]{\bf suspension length [\(\Delta\)]}, and (2) \mybox[fill=gray]{\bf the number of nested interruptions [\(|w|\)]}. 
Figure \ref{fig:E2-General} presents the eight independent (\(\imath\nu_{1-8}\)) variables of this study,
%(by conducting a comprehensive literature review on task switching and interruption analysis), 
the way we interpreted them in the course of our data analysis, their data collection method, as well as their corresponding literature references.
% Activation controls the speed and reliability of access to the memories and strength determines the degree to which we can reactivate old 

\begin{table*}
\vspace{-5mm}
  \caption*{}
 % \footnotesize
  \label{tab:freq}
  \begin{tabular}{p{17cm}}
    \toprule
  \\
  \begin{minipage}{.96\linewidth}
  \centering
\vspace{-5mm}
\includegraphics[scale=1.8]{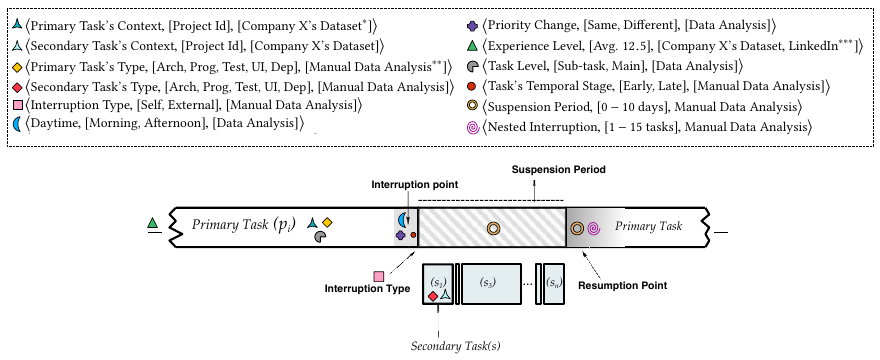}
\vspace{-5.5mm}
\captionof{figure}{\small An overview of task switching spectrum and independent variables of this study. Legend's symbols can be interpreted as \(\big<\)independent variable, \([\){potential values}\(]\), {data collection method}\(\big>\). [*]: We used the project ID provided in the dataset to distinguish different contexts. [**]: We used text mining and manual analysis on the metadata associated with each task to explore the type of the tasks under study. [***]: We used employee's LinkedIn account to extract required information about their work experience. \copyright ~Shakeri H.A , Karras, Schneider, Barker. }
\label{fig:E2-General}
\end{minipage}\\
\\

\underline{ Independent Variables (Task Switching Characteristics)  }\\
%\ding{192} {\it Task type:}\\
% \ding{193} {\it Task similarity:}\\

\ding{192} \emph{\bf \small Context Switching [CS=1, Different project] }(\(\imath\nu_1\)): switching the project along with task switching \cite{Pro1, Sky}. \\
\ding{193} {\bf  \small Type Difference [TD=1, Different type] }(\(\imath\nu_2\)): the type of the primary and the secondary tasks \cite{TofPS}. \\
\ding{194} {\bf \small Interruption Type [IT=1, Self] }(\(\imath\nu_3\)): \mybox[fill=gray]{Self-interruption} if the interruption initiated by the subject of the primary task; \mybox[fill=gray]{external-interruption} if it is motivated by some external events in the environment \cite{Mind, Theory2}. \\
\ding{195} {\bf \small Daytime [DT=1, Morning] }(\(\imath\nu_4\)): The time of the day that task switching occurs \cite{Behind}. All task switching and interruptions that were occurred between 11 am-1:30 pm (i.e. lunch time) were excluded from our analysis.\\
 \ding{196} {\bf \small Priority Change [PC=1, Same Priority] }(\(\imath\nu_5\)) \cite{TofPS}. \\
\ding{197} {\bf \small Experience Level [EL=1, More experience] }(\(\imath\nu_6\)): We recorded the
experience level of each of the included employees in our retrospective study from their LinkedIn account. The average professional software development experience of participants is 10.5 (range 4 to 25)~\cite{PS1, Reasoning}. \\
\ding{198} {\bf \small Task Level [TL=1, Sub-task]}(\(\imath\nu_7\)): the abstraction level of task \cite{Theory2}. We used ParentId column of the dataset to identify task levels. \\
\ding{199} {\bf \small Task Stage [TS=1, Late stage] }(\(\imath\nu_8\)): the completion state of the task \cite{Stage, Crazy}. We used temporal task logs and manually analyzed this dataset to identify the completion level of each task.\\

  \bottomrule
\end{tabular}
\vspace{-3mm}
\end{table*}

\subsection{Related Work}
Characterizing, managing, and theorizing multitasking and task switching have received increasing research attention from different disciplines such as psychology~\cite{Mind, MofGoals, Theory2}, human-computer interaction~\cite{Diary, Behind, Design2}, and management~\cite{Management}. 
In addition to the related work discussed in Section~\ref{sec:TC}, we focus on research related to multitasking and interruptions in the area of software engineering. Looking at multitasking and productivity, Vasilescu~\emph{et al.\/}~\cite{Sky} developed models and methods to measure the rate and breadth of developer's context-switching behaviour and studied how the switching behaviour affects developers' productivity. 
They found that a high rate of project switching per day results in a lower productivity, and developers who are involved in several projects generate more output than others.
Similarly, 
Meyer~\emph{et al.\/}~\cite{Pro1} conducted two studies to investigate software developers' personal perception of productivity and the factors which impact this productivity. 
The results of both studies revealed that developers perceive their day as productive when they complete many or big tasks without interruptions or context switches. 
However, they observed that participants performed significant task and activity switching while still feeling productive. 
In a follow-up study, Meyer \emph{et al.}~\cite{Meyer2017} found work habits and perceived productivity are related with each other and identified the time, user input, emails, and planned meetings as factors influencing productivity. Abad et al. \cite{SERIP, RE17,RENext17, Abad2018} recently conducted four studies to investigate the disruptiveness of task switching in software development projects as well as in requirements engineering tasks. They investigated the impact of interruption length on the duration of interrupted tasks and found that \emph{interruption length} of a specific task, regardless of the type of this task, does not influence its duration significantly. Moreover, they found that, compared to other types of development tasks, requirements engineering tasks are the most vulnerable tasks to task switching and interruptions.

In terms of the frequency of task switching and developers' productivity, 
Tregubov \emph{et al.}~\cite{Boehm} conducted a retrospective analysis and propose a way to evaluate the number of cross-project interruptions using self-reported develop work logs. 
The authors reported that developers who, on a typical day, are involved in two or more projects, spend 17\% of their development effort on cross-project interruptions. While the results of this work reveal a strong correlation between the number of projects and number of reported interruptions, it shows the correlation between the number of projects and effort spent on cross-project interruptions is relatively weak. 
Cruz~\emph{et al.}~\cite{Cruz2017} conducted a large-scale study to investigate the impact of work fragmentations on developers' productivity and found that work fragmentation is positively correlated with lower observed productivity for an entire session and longer suspension lengths strengthen this effect. 
Chong and Siino \cite{Chong2006} compared the behaviour and the disruptive impact of interruptions among paired and solo programmers. 
They found that various interruption characteristics such as time, type, and length of the interruptions as well as strategies for handling work interruptions are significantly different between paired and solo programmers. 
Similarly, \emph{Ko et al.}~\cite{Ko2007} conducted a study to understand information needs and the behaviour of task switching and interruptions in collocated software development teams. They found that coworkers are the most frequent source of information in software development teams which causes continual unavoidable task switching and interruptions due to an information need.

Our study confirms some of these results such as the negative impact of task switching on developers' productivity as well as multitasking challenges facing software development teams. 
Our study extends previous research in the following ways: (1) we model and investigate a comprehensive set of interruption characteristics including task-specific and context-specific factors and study the impact of these factors on task interruptions in various types of software development tasks; (2) we provide a comparison between various development tasks (i.e. programming, testing, architecture design, interface design, and deployment) in terms of their vulnerability to interruptions and task switching.
The comprehensiveness of this work in terms of the size of our
datasets and the number of dependent and independent variables further builds on these past contributions.

%=================================================================================
 
\section{Methods}
                                                          
%=================================================================================
To achieve our study goals we followed a mixed methods approach including: (1) a longitudinal data analysis on 4,910 recorded tasks of 17 professional software developers, and (2) a user survey with 132 software practitioners to complement the quantitative results with developer perception on task switching and interruptions. 
%(+)(+)(+)(+)(+)(+)(+)(+)(+)(+)(+)(+)(+)(+)(+)(+)(+)(+)(+)(+)(+)(+)(+)(+)(+)(+)(+)

\subsection{Study 1: Retrospective Analysis}
To gain a broad view of how disruptive task switching and interruptions can be varied by interruption characteristics, we conduct a longitudinal, retrospective study of 4,910 recorded tasks of 17 professional software developers. 
During the 1.6 years of this study, we developed and tested our conceptual framework (e.g. dependent, independent, and confounding variables) through two exploratory studies. 
The first study was conducted on 7,770 recorded tasks of 10 employees to ensure dataset quality and to identify potential confounding variables, such as interruption source and type, experience level, and task stage. 
The second study explores the impact of various interruption characteristics on the disruptiveness of a very specific type of software development tasks and helped to garner additional insights into the problem of task switching in software development teams to better formulate the research's conceptual framework \cite{SERIP, RE17,RENext17}.
We conduct this study in collaboration with Arcurve~\footnote{www.arcurve.com}, a large Calgary independent software services company. 
The datasets required for these studies
were collected from Arcurves's task-based bug tracking and
project management tool (i.e. Fogbugz~\footnote{http://www.fogcreek.com/fogbugz}). 
For each employee, we recorded 100 interruptions giving us 1700 recorded interruptions~\footnote{The data extraction form and a sample dataset collected for one employee are available http://wcm.ucalgary.ca/zshakeri/projects}.

\subsection{Study 2: User Survey}

To garner additional qualitative insights into developers' perception of task switching and interruptions, we use a survey.
We sent an online survey to \(800\) professional software developers working at companies of various sizes (e.g. Microsoft, Tableau Software, Ericsson, Bosch, and Cisco). The survey included 30 question using multiple choice, Likert scale, and open-ended questions. 
We asked participants about their job roles, development experience in general, their perception of task switching and productivity and the interruption factors which influence their productivity. 
We received 132 complete responses (17\% response rate). Of all 132 participants, 90 (68\%) listed their job as a programmer, 18 (14\%) as a software architect, 16 (12\%) as a tester, 5 (4\%) as project manager and 3 (2\%) as requirements engineer. The average professional software development experience per participant was 11.3 years (median: 8; range 3 to 40). 
The majority (99 or 75\%) reported the size of their company (i.e. \(s\)= number of employees) \(s\geq1000\), 11 (8\%): \(100\leq s<1000\); 7 (5\%): \(50\leq s<100\), and 8 (6\%): \(s<50\). 
As an incentive, survey respondents were given the option of being entered into a raffle to win one of the \$50US Amazon gift cards.
%~\footnote{This study received ethics approval from the the Ethics Board of the $X$ University}.

%=================================================================================

%=================================================================================

\begin{figure}
\includegraphics[scale=0.45]{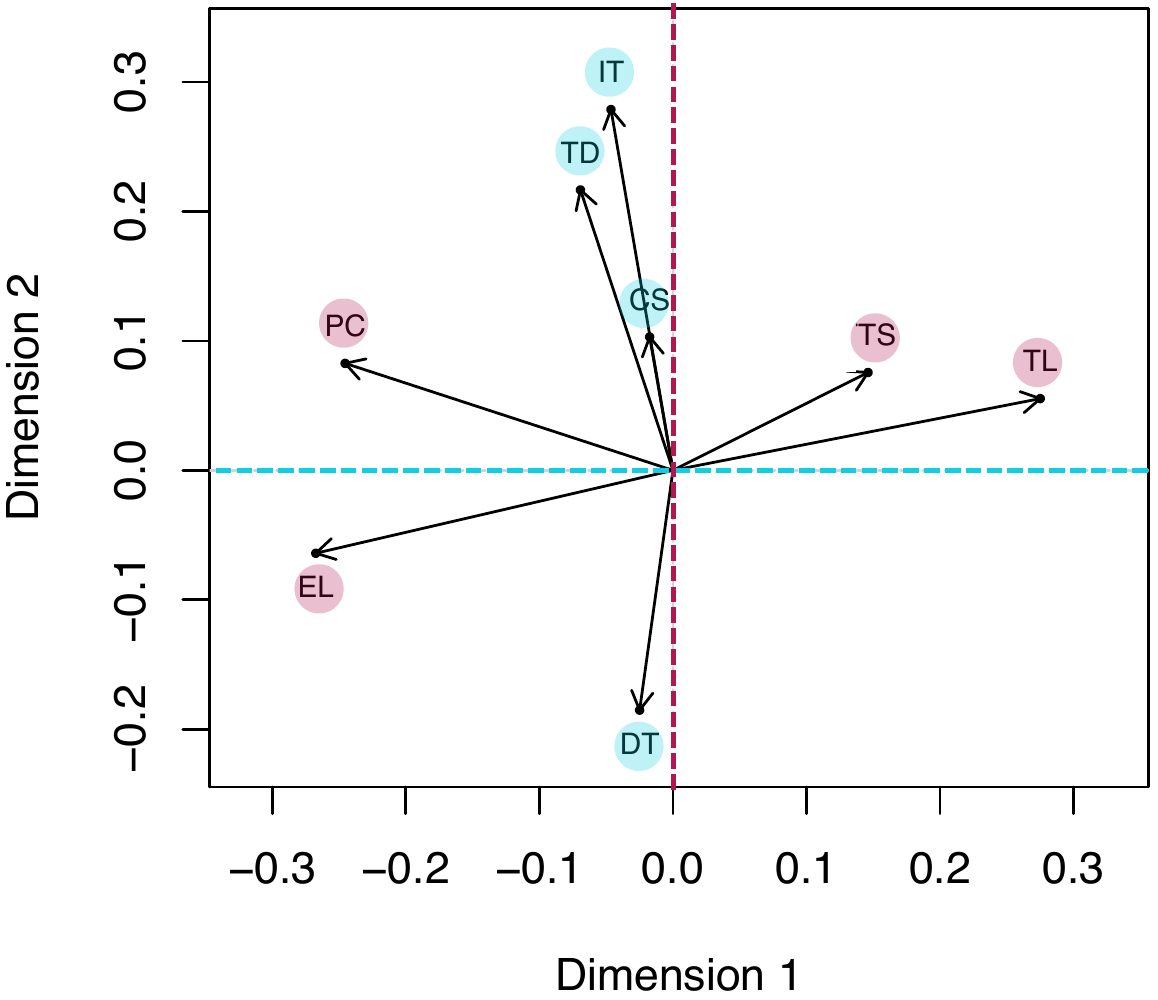}
\vspace{-3mm}
\caption{Loading plot for (\(\imath\nu_{1-8}\)) binary data}
\label{fig:Load}
\vspace{-5mm}
\end{figure}
%(+)(+)(+)(+)(+)(+)(+)(+)(+)(+)(+)(+)(+)(+)(+)(+)(+)(+)(+)(+)(+)(+)(+)(+)(+)(+)(+)

\subsection{Conceptual Framework}
The conceptual framework for our study draws from several lines of research and theory including multitasking studies~\cite{Sky}, the Memory of Goals \cite{MofGoals} and multitasking theories~\cite{Theory2}, and studies on developers' productivity and task management~\cite{Pro1, Meyer2017}.  
Recall Section~\ref{sec:TC} discusses eight independent (\(\imath\nu_{1-8}\)) and two dependent variables (\(\Delta, |w|\)) that are the major constructs of our study.  
To help interpret the results more easily, we apply \emph{homogeneity analysis} (i.e.  Multiple Correspondence Analysis (MCA)~\cite{MCA}) on \(\imath\nu_{1-8}\) to explore and summarize the underlying variable structure. 
As we recorded all of our independent variables in binary format, we used the \emph{non-linear Principal Component Analysis (PCA)} approach, a multivariate method for categorical data. To implement this approach, we used the \texttt{homals} function of package \texttt{\bf homals} \footnote{https://cran.r-project.org/web/packages/homals/homals.pdf} in {\bf R}. 
The loading plot presented in Figure \ref{fig:Load} helps identify variables that most contribute to each dimension. The loading scores of variables in each dimension are used as coordinates. 
The distance from each point (i.e. variable) to the abscissa (i.e. Dimension 1) or the ordinate (i.e. Dimension 2) gives a measure of the contribution of the point to each dimension. The greater the perpendicular distance from each point to an axis, the stronger the contribution of that point to the corresponding dimension \cite{MCA}. 
As illustrated in Figure \ref{fig:Load}, Dimension 1 has high loadings on \(\imath\nu_{1-4}\) (i.e. CS, TD, IT,  and DT) and describes {\bf context-specific characteristics} such as the context, type, and source of the task switching. 
Likewise, variables \(\imath\nu_{5-8}\) (i.e. PC, EL, TL, and TS)  contribute to Dimension 2, which describes {\bf task-specific characteristics} such as the abstraction level and the priority of the task as well as the required knowledge for performing the task. 
In the rest of this paper, we use these two dimensions for reporting and interpreting the results.

%(+)(+)(+)(+)(+)(+)(+)(+)(+)(+)(+)(+)(+)(+)(+)(+)(+)(+)(+)(+)(+)(+)(+)(+)(+)(+)(+)

\subsection{Research Questions (RQs)}
We formulated the following research questions:
\begin{description}
\item [RQ1- Task-specific Vulnerability:] How do various interruption characteristics impact the vulnerability of programming, testing, architecture design, UI design, and deployment tasks? 
\item [RQ2- Comparative Vulnerability:] Which types of development tasks are more vulnerable to task switching/interruptions?
\item [RQ3- Two-way Impact:] How does the interaction between various interruption characteristics (\(\imath\nu_{1-8}\)) influence the vulnerability of development tasks to interruptions? 
\end{description}

%In this paper, we deploy both nonparametric statistics and qualitative text analysis to address these RQs.
%(+)(+)(+)(+)(+)(+)(+)(+)(+)(+)(+)(+)(+)(+)(+)(+)(+)(+)(+)(+)(+)(+)(+)(+)(+)(+)(+)

\subsection{Data Analysis}

To test for the impact of disruptiveness factors and the difference between various task types (RQs 1-2), we use the non-parametric Kruskal-Wallis and Kruskal-Wallis posthoc tests, respectively. 
To determine the statistical significance we use the \emph{p-values} (\(\leq\) 0.05), and report as significant, differences at 95\% confidence interval, which we use to compare the disruptiveness of interruptions among different task types. 
Additionally, to check the correlation between participants' responses to survey questions, we use Spearman's rank test and define \(|\rho|\geq 0.50\) as a strong correlation coefficient. 
To model the cross-factor impact of disruptiveness factors (RQ3) we use the Scheirer-Ray-Hare (SRH) test, a non-parametric two-way ANOVA and an extension of the Kruskal-Wallis test. As a high correlation between predictor variables impact the statistical tests of predictors individually, we first applied Phi coefficient tests to statistically test the correlation between all of the independent variables for each of programming, testing, architecture/UI design, and deployment task types. 
For all correlated factors, we only use the two-way component of SRH tests and to statistically test the significant impact of individual disruptiveness factors on each task type, we applied the Kruskal-Wallis posthoc tests. To analyze the open-ended questions of the survey, we use a modified version of the grounded theory method~\cite{GTheory}, as a qualitative text analysis method, and use the Saturate App~\footnote{
www.saturateapp.com/} tool to code the survey responses.

%(+)(+)(+)(+)(+)(+)(+)(+)(+)(+)(+)(+)(+)(+)(+)(+)(+)(+)(+)(+)(+)(+)(+)(+)(+)(+)(+)

\section{Results}

\begin{table}
\centering
\footnotesize
\caption{\small Top 5 reasons for self-interruptions (\(\%(\#)\) represents percentage(number) of survey participants)}
\label{tab:Self}
\vspace{-3mm}
\begin{tabular}{p{6.8cm} p{0.8 cm}}\hline
{\cellcolor{gray}{ Reasons for self-interruptions/task switchings }}&\cellcolor{gray}\(\%(\#)\)\\\hline
Being blocked on a task (e.g. tool obstacles, technical issues)& 37 (30\%) \\
\cellcolor{G1}Getting sidetracked to other tasks (e.g. remembering other tasks, concentration lapse) &\cellcolor{G1} 28 (23\%)\\
Planning issues and priority changes (e.g. tasks with near due dates, short term deadlines)& 23 (19\%)\\
\cellcolor{G1}A need for more information/ technical knowledge (e.g. lack of documentation, waiting for feedback)& \cellcolor{G1}20 (16\%)\\ 
Getting bored with the task (e.g. complex and lengthy tasks) & 15 (12\%)\\\hline
\end{tabular}
\vspace{-3mm}
\end{table}
%000000000000000000000000000000000000000000000000000000
{\bf Practitioners' Perceptions of Task Switching and Interruptions:} When asked about whether participants consider task switching a type of interruption, 107 (81\%) stated that they consider task switching a specific type of interruptions because there is always some ramp-up time when switching between tasks as described by one participant's comment: \emph{``Saying that task switching is not an interruption sounds like multitasking is possible. It is not possible and changing the task will interrupt the other task every time and it takes approximately 5-20 minutes to get into the flow state on the task at hand every time there is a switch''}. 
We asked survey participants to list the main reasons that would make them have unplanned task switching. 
We iterated through the responses using the grounded theory approach ~\cite{GTheory}. 
Recall from Table \ref{tab:Self}, getting blocked or getting sidetracked to other tasks, planning issues, a need for more information, and boredom are the most common written responses to this question.

%_))))))))))))))))))))))))))))))))))))))))))))))))))))))))))

\bgroup

\def\arraystretch{0.75}

\begin {table}
\vspace{-2mm}
\scriptsize
\centering
\caption {\small RQ1- Impact of interruption characteristics on different task types. [{\tiny\ding{108}}] represents characteristics that make interruptions significantly more disruptive (based on 95\% confidence analysis).}
\vspace{-3mm}
\label{tab:RQ2}
\setlength{\tabcolsep}{.15em}
\begin{tabular} {|p{.2cm}|p{1.75cm}!{\color{black}\vrule}p{0.5cm}p{0.41cm}!{\color{black}\vrule}p{0.5cm}p{0.5cm}!{\color{black}\vrule}p{0.45cm}p{0.5cm}!{\color{black}\vrule}p{0.41cm}p{0.35cm}!{\color{black}\vrule}p{0.41cm}p{0.41cm}!{\color{black}\vrule}} \hline

&\textcolor{white}{.........}{\bf Pairs}&\multicolumn{2}{c|}{{Arch}}&\multicolumn{2}{c|}{Prog}&\multicolumn{2}{c|}{Test}&\multicolumn{2}{c|}{{UI}}&\multicolumn{2}{c|}{{Dep}}\\\cline{3-12}
{}&{}&\(\Delta\)&\(|w|\)& \(\Delta\)&\(|w|\)&  \(\Delta\)&\(|w|\)&  \(\Delta\)&\(|w|\)&  \(\Delta\)&\(|w|\)\\\hline
\multirow{12}{*}[-2ex]{\rotatebox[origin=c]{90}{Context-specific Factors}}&{Kruskal Wallis}&\cellcolor{gray}0.01&0.06&\cellcolor{gray}0.002&\cellcolor{gray}0.03&0.3&0.06&0.1&0.6&\cellcolor{gray}0.03&0.08\\\arrayrulecolor{black}\cline{2-12}
{}&{\bf same project}&&&&&&&&&&\\
{}&{\bf \cellcolor{lb}diff project}&\multicolumn{1}{c}{\cellcolor{lb} \ding{108}}&&\multicolumn{1}{c}{\cellcolor{lb} \ding{108}}&\multicolumn{1}{c!{\color{black}\vrule}}{\cellcolor{lb} \ding{108}}&&&&&\multicolumn{1}{c}{\cellcolor{lb} \ding{108}}&\\\arrayrulecolor{black}\cline{2-12}

&{Kruskal Wallis}&0.5&0.1&0.06&\cellcolor{gray}2e-9&\cellcolor{gray}{5e-6}&\cellcolor{gray}5e-11&0.8&0.7&0.2&\cellcolor{gray}0.01\\\arrayrulecolor{black}\cline{2-12}
{}&{\bf diff type} &&&&&&&&&&\\
{}&{\bf \cellcolor{lb}same type} &&&&\multicolumn{1}{c!{\color{black}\vrule}}{\cellcolor{lb} \ding{108}}&\multicolumn{1}{c}{\cellcolor{lb} \ding{108}}&\multicolumn{1}{c!{\color{black}\vrule}}{\cellcolor{lb} \ding{108}}&&&&\multicolumn{1}{c!{\color{black}\vrule}}{\cellcolor{lb} \ding{108}}\\\arrayrulecolor{black}\cline{2-12}

{}&{Kruskal Wallis}&\cellcolor{gray}0.04&0.2&\cellcolor{gray}2e-9&\cellcolor{gray}2e-8&\cellcolor{gray}2e-9&\cellcolor{gray}3e-9&\cellcolor{gray}0.02&0.5&\cellcolor{gray}0.01&\cellcolor{gray}0.004\\\arrayrulecolor{black}\cline{2-12}
{}&{\bf \cellcolor{lb}self-Interruption}&\multicolumn{1}{c}{\cellcolor{lb} \ding{108}}&&\multicolumn{1}{c}{\cellcolor{lb} \ding{108}}&\multicolumn{1}{c!{\color{black}\vrule}}{\cellcolor{lb} \ding{108}}&\multicolumn{1}{c}{\cellcolor{lb} \ding{108}}&\multicolumn{1}{c!{\color{black}\vrule}}{\cellcolor{lb} \ding{108}}&\multicolumn{1}{c}{\cellcolor{lb} \ding{108}}&&\multicolumn{1}{c}{ \cellcolor{lb}\ding{108}}&\multicolumn{1}{c!{\color{black}\vrule}}{\cellcolor{lb} \ding{108}}\\
{}&{\bf ext-Interruption} &&&&&&&&&&\\\arrayrulecolor{black}\cline{2-12}

&{Kruskal Wallis}&0.5&0.1&\cellcolor{gray}0.001&\cellcolor{gray}1e-5&\cellcolor{gray}0.01&\cellcolor{gray}0.03&\cellcolor{gray}4e-8&0.2&0.6&0.9\\\arrayrulecolor{black}\cline{2-12}
{}&{\bf morning}&&&\multicolumn{1}{c}{\cellcolor{lb} \ding{108}}&&&&&&&\\
{}&{\bf \cellcolor{lb}afternoon}&&&&\multicolumn{1}{c!{\color{black}\vrule}}{\cellcolor{lb} \ding{108}}&\multicolumn{1}{c}{\cellcolor{lb} \ding{108}}&\multicolumn{1}{c!{\color{black}\vrule}}{\cellcolor{lb} \ding{108}}&{\cellcolor{lb} \ding{108}}&&&\\\arrayrulecolor{black}\cline{2-12}
\hline
\hline

\multirow{12}{*}[-2ex]{\rotatebox[origin=c]{90}{Task-specific Factors}}&{Kruskal Wallis}&0.2&0.4&0.5&\cellcolor{gray}1e-7&0.6&\cellcolor{gray}0.01&\cellcolor{gray}0.02&0.8&0.9&0.8\\\arrayrulecolor{black}\cline{2-12}
{}&{\bf \cellcolor{lb}same priority} &&&&\multicolumn{1}{c!{\color{black}\vrule}}{\cellcolor{lb} \ding{108}}&&\multicolumn{1}{c!{\color{black}\vrule}}{\cellcolor{lb} \ding{108}}&{\cellcolor{lb} \ding{108}}&&&\\
{}&{\bf diff priority} &&&&&&&&&&\\\arrayrulecolor{black}\cline{2-12}

&{Kruskal Wallis}&0.2&0.8&0.6&0.2&\cellcolor{gray}0.01&0.2&0.2&0.06&\cellcolor{gray}0.01&\cellcolor{gray}0.03\\\arrayrulecolor{black}\cline{2-12}
{}&{\bf \cellcolor{lb}less experience} &&&&&\multicolumn{1}{c}{\cellcolor{lb} \ding{108}}&&&&\multicolumn{1}{c}{\cellcolor{lb} \ding{108}}&\multicolumn{1}{c!{\color{black}\vrule}}{\cellcolor{lb} \ding{108}}\\
{}&{\bf more experience} &&&&&&&&&&\\\arrayrulecolor{black}\cline{2-12}

{}&{Kruskal Wallis}&\cellcolor{gray}0.002&0.06&\cellcolor{gray}0.03&\cellcolor{gray}0.01&0.4&0.2&\cellcolor{gray}0.01&0.1&\cellcolor{gray}0.01&\cellcolor{gray}0.03\\\arrayrulecolor{black}\cline{2-12}
{}&{\bf sub-task}&&&&&&&&&&\\
{}&{\bf \cellcolor{lb}main task}&\multicolumn{1}{c}{\cellcolor{lb} \ding{108}}&&\multicolumn{1}{c}{\cellcolor{lb} \ding{108}}&\multicolumn{1}{c!{\color{black}\vrule}}{\cellcolor{lb} \ding{108}}&&&\multicolumn{1}{c}{\cellcolor{lb} \ding{108}}&&\multicolumn{1}{c}{\cellcolor{lb} \ding{108}}&\multicolumn{1}{c!{\color{black}\vrule}}{{\cellcolor{lb} \ding{108}}}\\\arrayrulecolor{black}\cline{2-12}

{}&{Kruskal Wallis}&0.7&0.5&0.3&0.3&0.1&0.2&0.1&0.5&0.1&\cellcolor{gray}0.04\\\arrayrulecolor{black}\cline{2-12}
{}&{\bf early stage}&&&&&&&&&&\\
{}&{\bf \cellcolor{lb}late stage} &&&&&&&&&&\multicolumn{1}{c!{\color{black}\vrule}}{\cellcolor{lb} \ding{108}}\\\arrayrulecolor{black}\hline
\end{tabular}
\vspace{-5mm}
\end{table}
\egroup

\subsection{RQ1- Task-specific Vulnerability}
We follow a template and posed 80 null hypotheses to explore factors that may explain the disruptiveness of interruptions in various types of software development tasks:
{\bf \(H_0 =\) Interruption characteristic \(\imath \nu_i\) does not impact the \(\Delta\) and/or \(|w|\) of task switchings in task \(\langle T\rangle\). } Where $\Delta$ denotes the suspension period, \(|w|\) the length of nested task switching, and \(\langle T\rangle\) denotes the task type. As illustrated in Table \ref{tab:RQ2}, of 34 (43\%) rejected tests, 21 (62\%) are related to \emph{contextual} factors, and 13 (38\%)  are related to \emph{task-specific} factors. 
This implies that, compared to task-specific factors, contextual factors (e.g. context switching and interruption type) are more potent determinants of task switching disruptiveness in software development tasks. 

{\bf Finding \(_{\bf 1-1}\):} The interruption type (i.e. self/external) significantly impacts at least one disruptiveness factor for {\bf all} of the task types under study. As illustrated in Table \ref{tab:RQ2}, {\bf self-interruptions} make task switching and interruptions more disruptive by negatively impacting the length of the suspension period and the number of nested interruptions. 
Task level (i.e. sub-task/main) comes next, with significant impact on four task types (i.e. architecture, programming, UI, and deployment). 
Context switching and type switching each negatively impacts three task types. 
 
{\bf Finding \(_{\bf 1-2}\):} Priority change, daytime, and type difference are characteristics that significantly impact both programming and testing tasks' interruptions. 
Looking at Table \ref{tab:RQ2}, the 95\% confidence analysis shows that afternoon interruption or switching to another task with the same priority, or a different type makes programming/testing task interruptions more vulnerable. 
Moreover, while context switching does not significantly impact the vulnerability of testing and UI tasks to interruptions, switching to a different project negatively impacts the \(\Delta\) of architectural, programming, and deployment tasks. 
  
{\bf Finding \(_{\bf 1-3}\):} Following the results of the Kruskal-Wallis tests, only testing interruptions are significantly impacted by the \emph{experience level} (\(p\)=0.01). 
Table \ref{tab:RQ2} shows less experienced testers are more vulnerable to interruptions than experienced ones.
Likewise, \emph{task stage} impacts only one task type (i.e. deployment tasks). 

{\bf Discussion \(_{\bf 1-1}\):} Although our  analysis revealed the statistically significant negative impact of self-interruptions on the vulnerability of \emph{all} development tasks, 107 (81\%) participants stated \emph{external-interruptions} are more disruptive than self-interruptions. 
When asked with an open-ended question about the impact of interruption type on their productivity, most of the participants who selected external interruptions, stated  external interruptions are unexpected and are not in their control so are more disruptive. 
They believed they cannot control the timing of these interruptions which subsequently negatively impacts their performance when they resume the interrupted task, as evidenced in the following quote from one of the participants: \emph{``I tend not to have control over these interruptions and thus I need to follow what they are saying and find a way to make what they are saying happen, and this causes me to become very involved with that one thing which takes time''}. 
However, the results of two recent studies conducted by Katidioti \emph{et al.}~\cite{SelfExternal} comparing the disruptiveness of self and external interruptions support the results of our quantitative analysis and reveal that external-interruptions are less disruptive than self-interruptions. 
Similarly, a recent study by Adler and Benbunan-Fich~\cite{SelfExternal2} shows that more self-interruptions result in lower accuracy in resumed tasks which causes performance difficulties and consequently sub-optimal results.
Another participant of our survey who selected self-interruptions as more disruptive stated that: \emph{``External interruptions are disruptive, but do not necessarily add more items to my cognitive stack. Internal interruptions are always caused by me having (or perceiving myself to have) too many tasks to solve''}. 
\begin{figure}
\centering
\subfloat[Portion of external interruptions]{\includegraphics[scale=0.39]{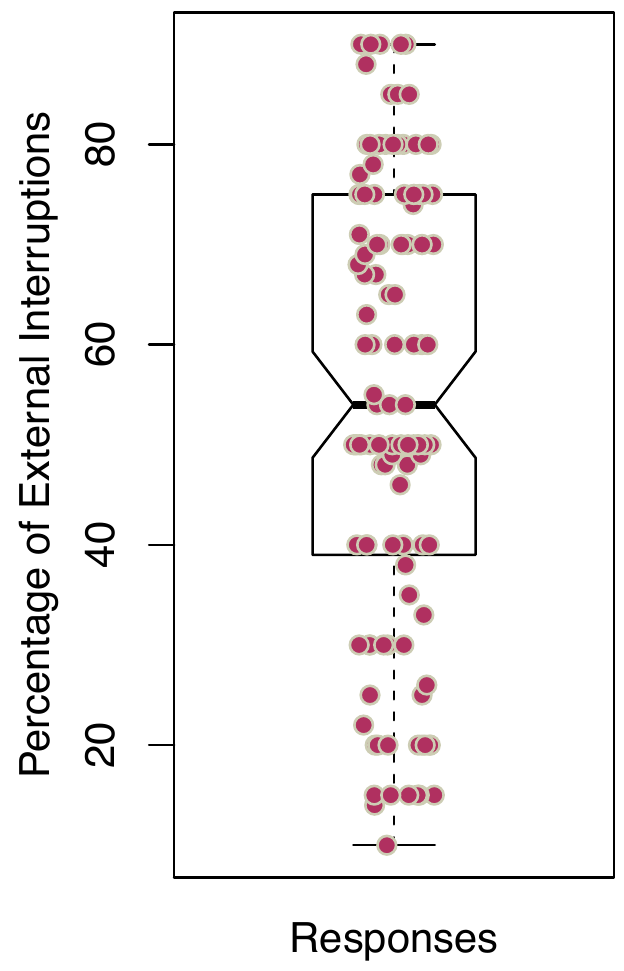}}\hspace{1em}
\subfloat[Spearman's ranked correlations]{\includegraphics[scale=0.37]{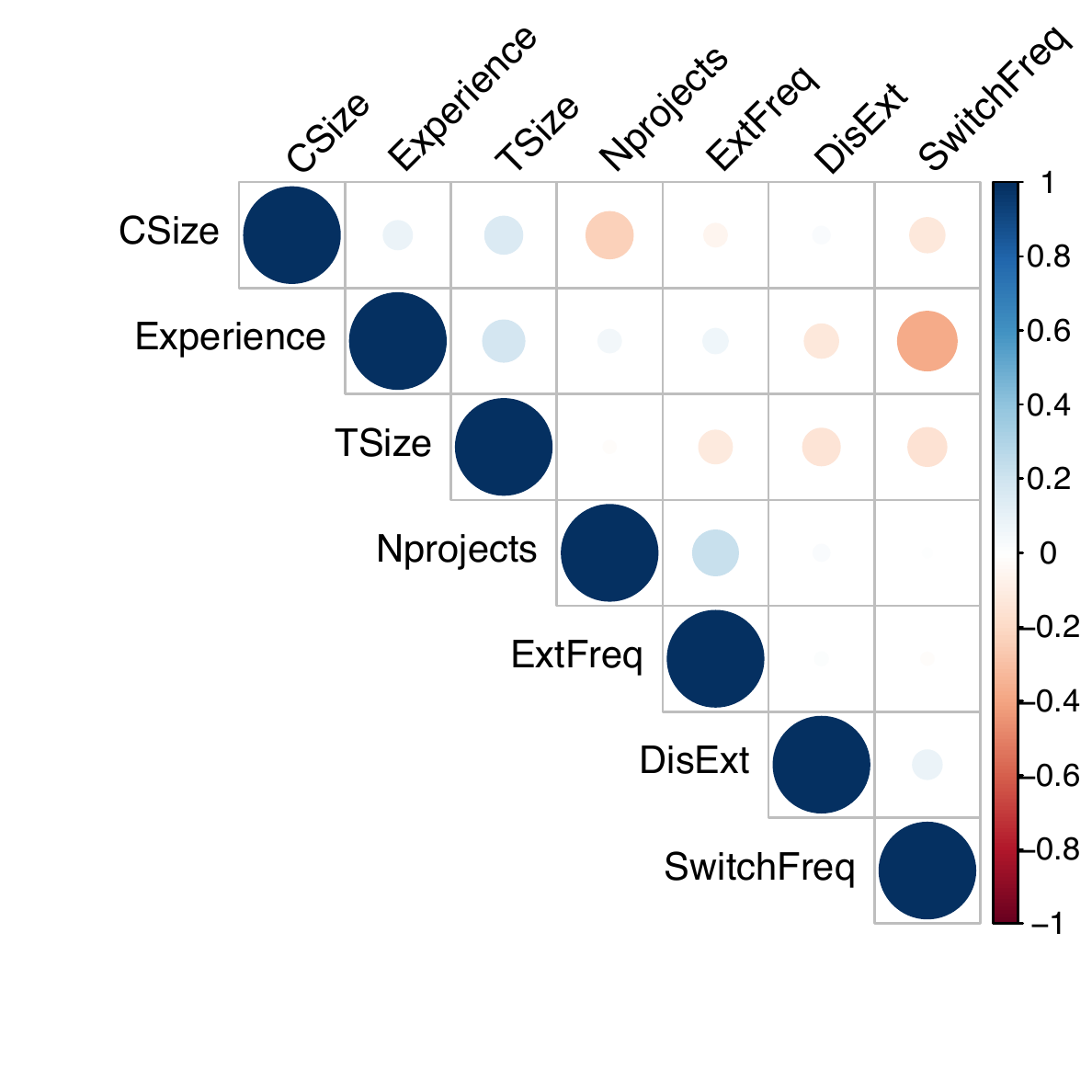}}
\vspace{-3.5mm}
\caption{\small Perceived frequency and disruptiveness of external interruptions and correlation analysis. [CSize=Company size, TSize=Team Size, NProject= \# of projects that participants are involved in, ExFreq= the frequency of external interruptions, DisExt= Disruptiveness of external interruptions, SwitchFreq= the frequency of task switching]}
\label{fig:Normal}
\vspace{-3mm}
\end{figure}

We speculate that the difference between our survey results and the results of our retrospective analysis and existing theoretical and practical evidence could be due to the high frequency of external interruptions in software development environments. 
We asked survey participants to, on a scale from 1 to 100, rate what portion of their task switching and interruptions in a day are triggered by an external event. It can be seen from Figure \ref{fig:Normal}a that responses given to this question are slightly skewed to the left which implies that frequencies are more towards the higher side, with mean (and median) values of 54\% (range 10-90\%). 
Moreover, we further investigated the association between the disruptiveness and frequency of external interruptions reported by participants and other factors such as their company and team size as well as their experience level and the number of projects they contribute to on a typical day. 
Spearman's rank correlation tests (summarized in Figure \ref{fig:Normal}b) show the perceived frequency and the disruptiveness of external interruptions do not correlate with their team size, experience level, or the number of projects they are involved in (e.g. TSize-ExtFreq: rho= -0.12, \(p\)=0.2; TSize-DisExt:  rho= -0.15, \emph{p}=0.12).

\begin{figure}
\includegraphics[scale=0.36]{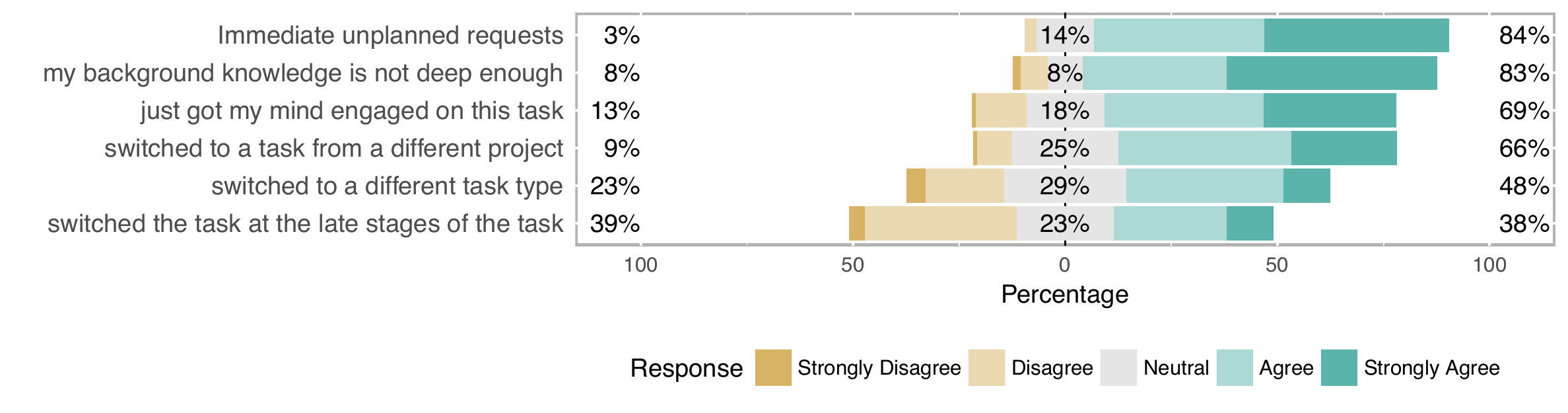}
\vspace{-5mm}
\caption{Perceived impact of interruption characteristics}
\vspace{-5mm}
\label{fig:Likert1}
\end{figure}

 {\bf Discussion \(_{\bf 1-2}\):} We asked survey respondents to rate the negative impact of context and type switching on a Likert-scale.
120 (91\%) and 102 (77\%) of the participants indicated neutrality or agreement about the negative impact of context switching and type changes, respectively (Figure \ref{fig:Likert1}). 
The participants predominantly stated that context switching requires a different mindset which places more demands on cognitive resources and makes task switching more disruptive: \emph{``while it depends on how much you have to remember about a specific task/project, context-switching can require more ramp-up because there's more context you have to bring back up''}. 
This finding is supported by existing literature~\cite{Meyer2017, Pro1, Boehm} evaluating the negative impact of context switching on work fragmentation and consequently on developers' productivity and quality of work produced. 

\begin {table*}
\scriptsize
\centering
\vspace{-3mm}
\caption {RQ2- Comparison between various development tasks concerning their vulnerability to interruption}
\vspace{-4mm}
\label{tab:RQ1}
\setlength\tabcolsep{3.9pt}
\begin{tabular} {|p{3.5cm}|p{0.55cm}p{0.6cm}|p{0.55cm}p{0.5cm}|p{0.6cm}p{0.63cm}|p{0.48cm}p{0.5cm}||p{0.55cm}p{0.55cm}|p{0.55cm}p{0.6cm}|p{0.45cm}p{0.45cm}|p{0.5cm}p{0.5cm}|} \hline
{}&\multicolumn{8}{c||}{{\bf Context-specific (Dimension 2)}}&\multicolumn{8}{c|}{{\bf Task-specific (Dimension 1)}}\\\cline{2-17}
\textcolor{white}{.........}{\bf Pairs}&\multicolumn{2}{c|}{{different context}}&\multicolumn{2}{c|}{different type}&\multicolumn{2}{c|}{interruption type}&\multicolumn{2}{c||}{{daytime}}&\multicolumn{2}{c|}{{priority change}}&\multicolumn{2}{c|}{{experience level}}&\multicolumn{2}{c|}{{task level}}&\multicolumn{2}{c|}{{temporal stage}}\\
{}&\multicolumn{2}{c|}{{\(\imath\nu_1 \) ({\it CS=1 }})}&\multicolumn{2}{c|}{{\( \imath\nu_2\) ({\it TD=1}) }}&\multicolumn{2}{c|}{{\(\imath\nu_3\) ({\it IT=1})}}&\multicolumn{2}{c||}{{\(\imath\nu_4\) ({\it DT=1})}}&\multicolumn{2}{c|}{{ \(\imath\nu_5\) ({\it PC=1})}}&\multicolumn{2}{c|}{{ \(\imath\nu_6\) ({\it EL=0})}}&\multicolumn{2}{c|}{{\(\imath\nu_7\) ({\it TL=1})}}&\multicolumn{2}{c|}{{\(\imath\nu_8\) ({\it TS=0})}}\\\cline{2-17}
{}&\(\Delta\)&\(|w|\)& \(\Delta\)&\(|w|\)&  \(\Delta\)&\(|w|\)&  \(\Delta\)&\(|w|\)&  \(\Delta\)&\(|w|\)&  \(\Delta\)&\(|w|\)&  \(\Delta\)&\(|w|\)&  \(\Delta\)&\(|w|\)\\\hline

{Kruskal-Wallis}&\cellcolor{light}{3e-4}& \cellcolor{light}{2e-4}& \cellcolor{light} 0.001& \cellcolor{light} 0.001& 0.4&  \cellcolor{light} 0.05&\cellcolor{light} 0.02&\cellcolor{light}{0.02}& \cellcolor{light} {2e-5}& \cellcolor{light}  4e-6&\cellcolor{light}2e-4&\cellcolor{light}0.003&   \cellcolor{light}4e-4&\cellcolor{light}1e-4&0.1&   \cellcolor{light}0.03\\\hline

{\bf Programming-Architecture}&\cellcolor{light}0.02*&\cellcolor{light}0.03*&0.2&0.2&0.2& \cellcolor{light}0.04&\cellcolor{light}0.03*& 0.1&\cellcolor{light}0.004*&\cellcolor{light}0.003*&0.8&\cellcolor{light}0.04*&\cellcolor{light}0.01&\cellcolor{light}0.05&0.3&0.2\\

{\bf Programming-Test}&\cellcolor{light}0.001&\cellcolor{light}0.001&\cellcolor{light}0.01&0.1& 0.4&0.9&0.6&0.4&\cellcolor{light}0.003&0.3&0.3&0.4&0.3&0.25&0.07&0.2\\

{\bf Programming-UI}&0.5&\cellcolor{light}0.03*&0.2&\cellcolor{light}0.05&0.3&\cellcolor{light}0.001&\cellcolor{light}0.04&\cellcolor{light}0.02&0.4&0.2&\cellcolor{light}0.04&\cellcolor{light}0.01&\cellcolor{light}0.04&\cellcolor{light}0.02&0.1&0.08\\

{\bf Programming-Deployment}&0.1&0.06&0.1&\cellcolor{light}0.01&0.3&0.2&0.9& \cellcolor{light}0.05&\cellcolor{light}0.02&\cellcolor{light}0.01&\cellcolor{light}0.01&\cellcolor{light}0.01&\cellcolor{light}0.01*&\cellcolor{light}0.02*&0.1&\cellcolor{light}0.01\\

{\bf Test-Architecture}&\cellcolor{light}0.05*&\cellcolor{light}0.03*&0.9&0.8&0.1&\cellcolor{light}0.04&0.07&0.3&0.3&0.6&0.7&0.4&\cellcolor{light}0.05&0.3&0.9&0.6\\

{\bf Test-UI}&0.2&\cellcolor{light}0.04*&0.9&\cellcolor{light}0.02*&0.2&\cellcolor{light}0.01&0.5&0.9&0.2&0.1&\cellcolor{light}0.01&\cellcolor{light}0.01&0.2&0.2&0.4&0.2\\

{\bf Test-Deployment}&0.2&\cellcolor{light}0.001&\cellcolor{light}0.01&\cellcolor{light}0.002&0.5&0.2&\cellcolor{light}0.02&\cellcolor{light}0.02&0.1&\cellcolor{light}0.03&0.1&\cellcolor{light}0.04&\cellcolor{light}0.001*&\cellcolor{light}0.002*&\cellcolor{light}0.02&\cellcolor{light}0.001\\

{\bf Architecture-UI}&0.9&0.9&0.9&0.7&0.9&0.5&0.07&0.3&0.07&0.1&0.1&0.2&0.4&0.8&0.6&0.5\\

{\bf Architecture-Deployment}&0.08&\cellcolor{light}0.02&\cellcolor{light}0.04&\cellcolor{light}0.001&0.2&\cellcolor{light}0.02&0.1&\cellcolor{light}0.01&0.4&\cellcolor{light}0.04*&0.1&\cellcolor{light}0.02&0.2&\cellcolor{light}0.02&\cellcolor{light}0.05&\cellcolor{light}0.003\\

{\bf Deployment-UI}&0.1&0.06&0.3&\cellcolor{light}0.01&0.2&\cellcolor{light}0.01&\cellcolor{light}0.001&\cellcolor{light}0.002&\cellcolor{light}0.02&\cellcolor{light}0.01&\cellcolor{light}0.001&\cellcolor{light}1e-4&0.1&0.6&\cellcolor{light}0.02&\cellcolor{light}0.001\\\hline
\multicolumn{17}{l|}{\it \scriptsize*: The p-value of the alternative value of the corresponding variable. }\\
\end{tabular}
\vspace{-3mm}
\end{table*}

{\bf Discussion \(_{\bf 1-3}\):} While our analysis shows a limited contribution of \emph{experience level} to the vulnerability of development tasks with interruptions, 110 (83\%) participants stated that task switching in situations where their background knowledge of performing a task is shallow or they are learning, negatively impacts their performance in the primary task: \emph{ ``[...] I don't have the most structured learning process, so sometimes the structure is not really clear in my head until I have explored a lot of it. If the structure is incomplete, then it's harder to remember, which means that any interruption will have a much worse impact on it than if I already knew the relevant area of code''}. 
Researchers have studied the effect of experience level on the cognitive load of tasks. 
Sweller~\cite{PS2} and Gregory \emph{et al.}~\cite{Reasoning} argue that experts have the ability to recognize the {\em problem state} from their previous experiences and accurately recall the information required for resuming their interrupted tasks. 
Conversely, novices are not able to memorize the problem state of their previous tasks and are forced to use their general problem-solving techniques to resume their interrupted tasks. 
 
Figure \ref{fig:Likert1} shows 91 (69\%) participants considered early stage interruptions as a factor that negatively affects their performance after resuming the primary task. 
The most common written response was that the early investment in a task is critical to building context about an issue and determining next steps when returning from an interruption.
This is particularly true in the early stages of a new project because  \emph{``early stage interruptions result in nearly a perfect storm of wasted time since the time I spent getting engaged had no pay-off''}. 
Moreover, only 50 (38\%) respondents considered late stage interruptions disruptive: \emph{``If the end is in sight, all the necessary work is laid out and is pretty easy to do without much thought. You've likely figured out the main points of the task if you are almost complete, at this point it's a matter of getting the work done and not figuring out how to do it''}. However, our retrospective analysis revealed that only deployment tasks are impacted by the temporal point of interruptions (\(p\)= 0.04), and this factor does not significantly impact the vulnerability of other development tasks to interruptions. Contrary to the survey results and the results of our repository analysis, several studies (e.g. \cite{TS, Stage}) investigated the impact of task stage on the cognitive cost of interruptions and found that middle or late stage interruptions cause longer suspension period (\(\Delta\)) and consequently decrease in performance and work quality. 
This difference raises questions about the cognitive cost of interruptions at different stages of a task and implies the need for a further investigation on this factor (i.e. TS).

{\bf Practitioner's corner \(_{\bf 1}\):} Considering the negative impact of self-interruptions on software developers' productivity (as discussed in Finding\(_{1-1}\) and Discussion\(_{1-1}\)), we recommend software developers minimize the frequency of their voluntary task switching. 
We also recommend that frequent context switching at either task type or project level negatively impacts programmers and testers' productivity by causing fragmented work and longer suspension length. 
Thus, since switching back and forth between different projects and task types decreases efficiency by forcing loading and unloading of context per switch, it might be more efficient if developers ask their questions from co-workers working on the same project/task type. 
Further, as stated by our survey participants, less experienced software developers find it harder to capture the context they were in before switching their primary task and they are most likely to need to backtrack further when they resume their interrupted tasks. 
Thus, software developers should ask their unplanned questions from co-workers who are more experienced in the topic related to their ongoing task. 
Consistent with other research~\cite{TS, Stage} and stated by 50 (38\%) participants of our survey, switching a task at late stages of the task causes more cognitive cost when recalling the task's context: \emph{``I have to rethink from the beginning to make sure that there was no mistake in the previous thoughts''}. 
However, as stated by one of the survey participants: \emph{``It depends more on complexity at the stage versus which stage in general. I have found it quite easy to resume later stage tasks if they are not complex. A lot of software development tasks are complex though so it could tend to be harder''}.
These apparent conflicts suggest additional research on this factor is required.
%many confounds should be considered to better understand the impact of this factor on software development tasks' interruptions. 	

\begin{figure*}
\vspace{-3mm}
\centering
\vspace{-1mm}
%\subfloat[]{\includegraphics[scale=0.7]{Figures/E1/DP-DevTest}}

%\subfloat[]{\includegraphics[scale=0.7]{Figures/E1/TK-Dev}}
\subfloat[]{\includegraphics[scale=0.125]{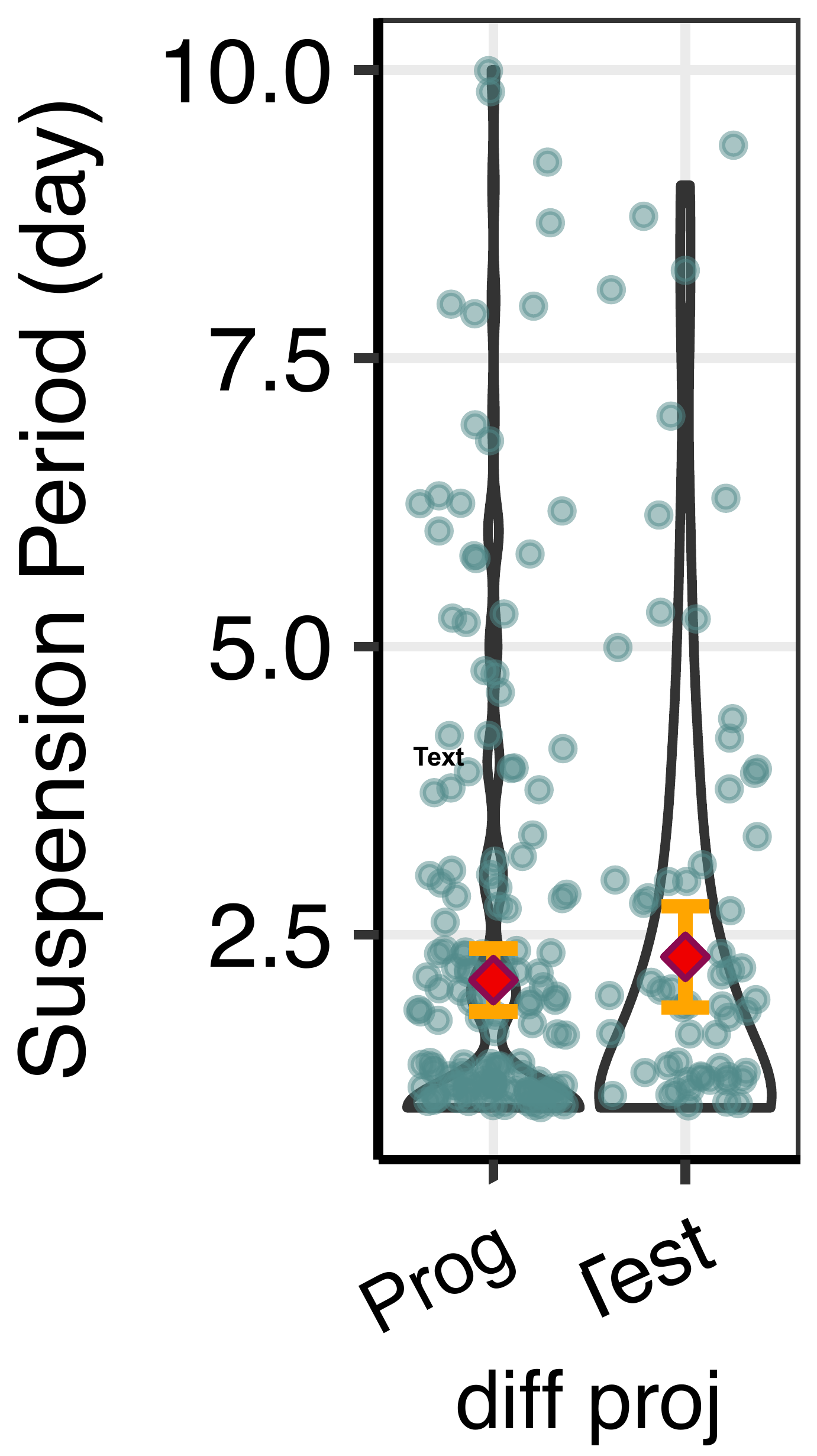}}
\subfloat[]{\includegraphics[scale=0.125]{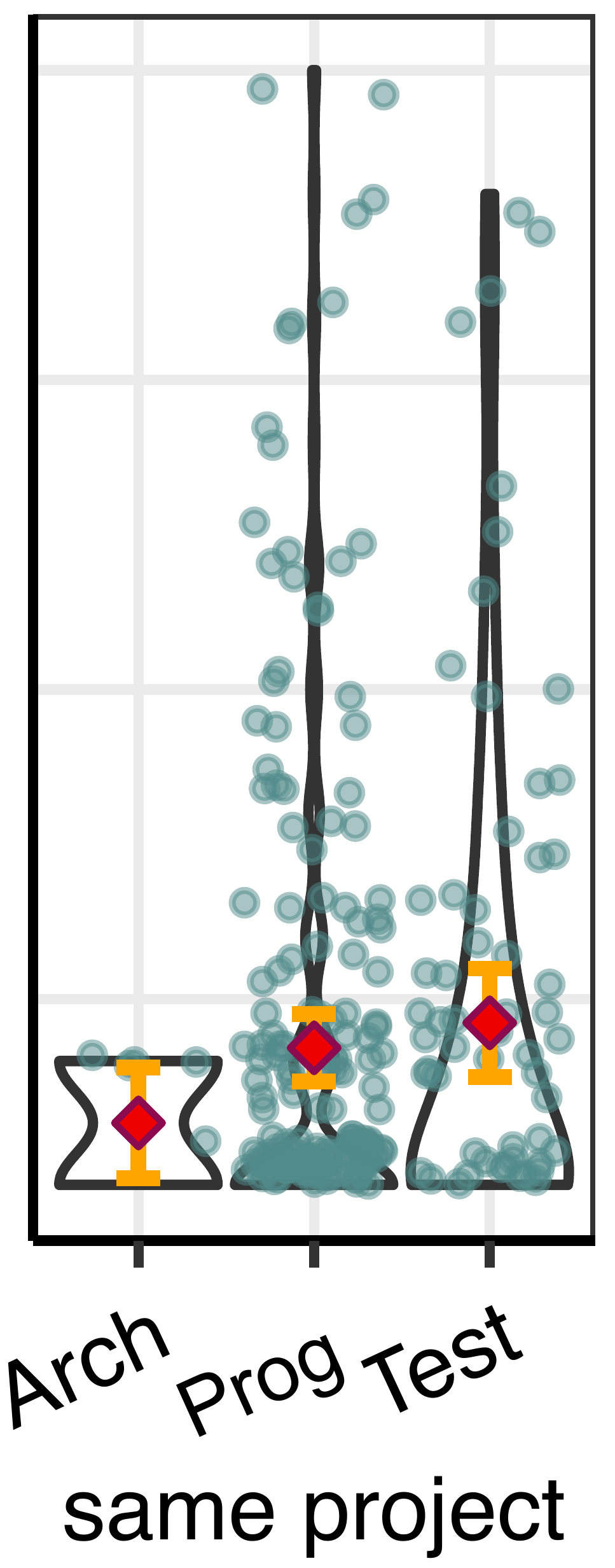}}
\subfloat[]{\includegraphics[scale=0.125]{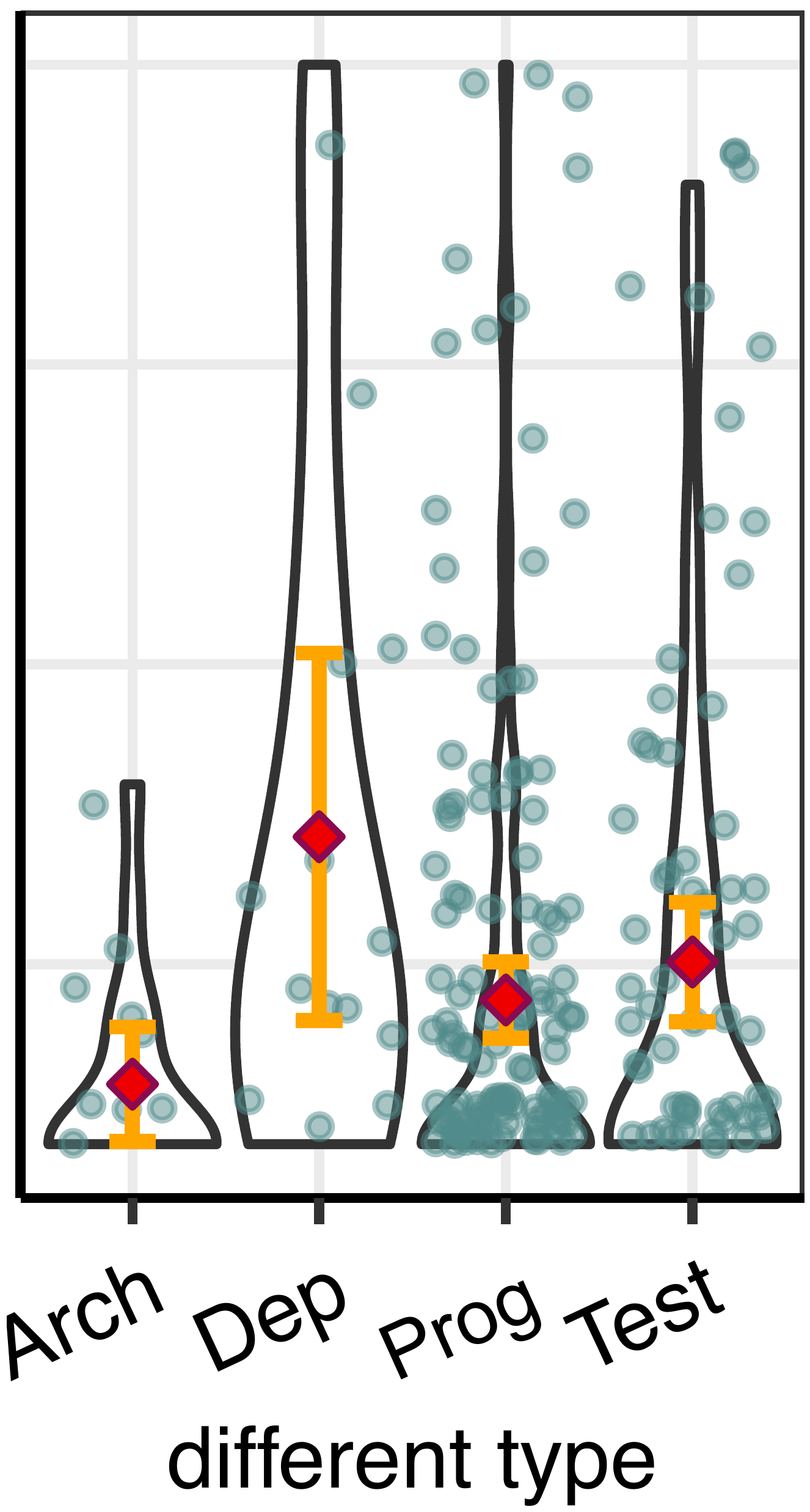}}
\subfloat[]{\includegraphics[scale=0.125]{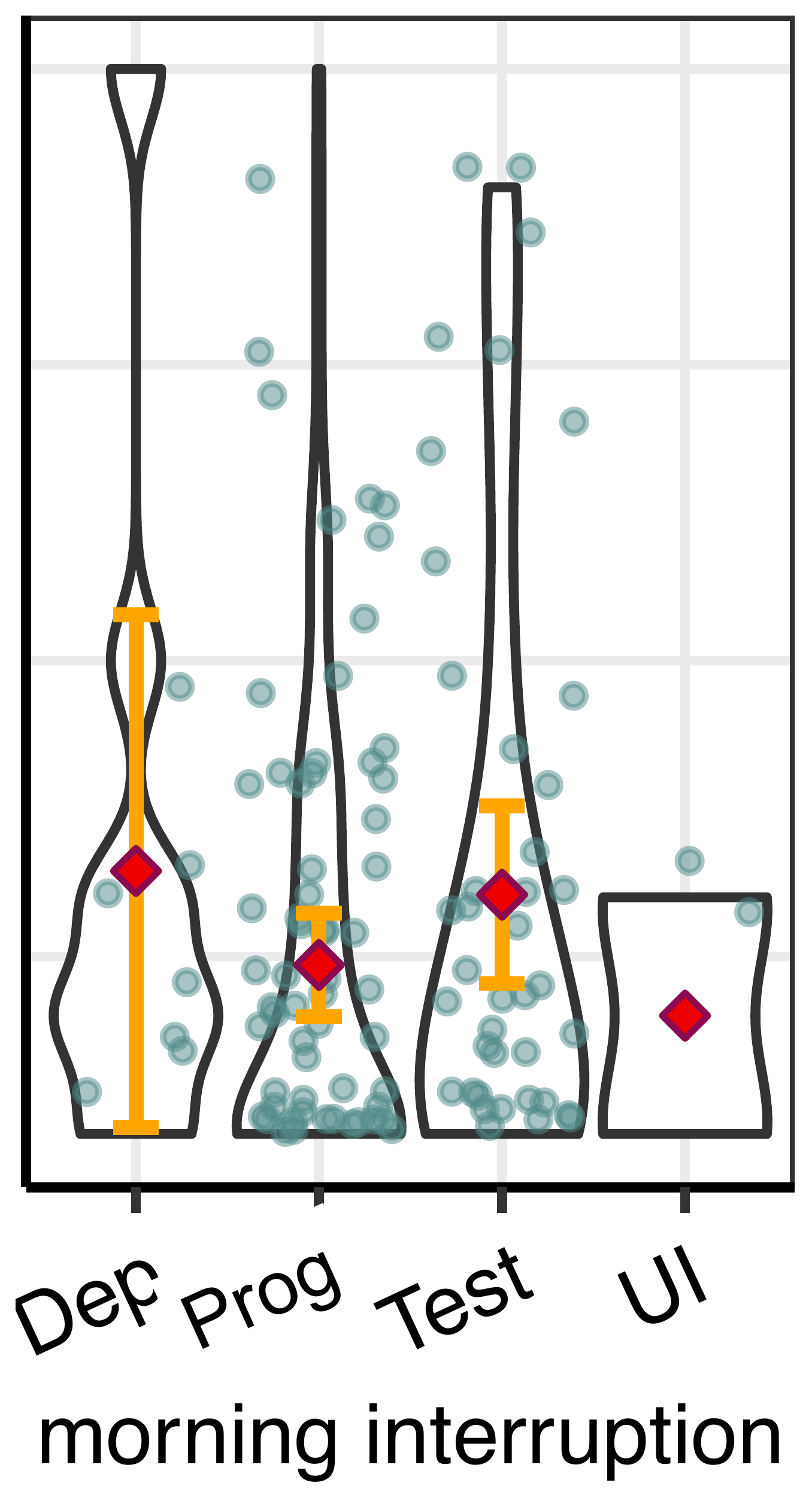}}
\subfloat[]{\includegraphics[scale=0.125]{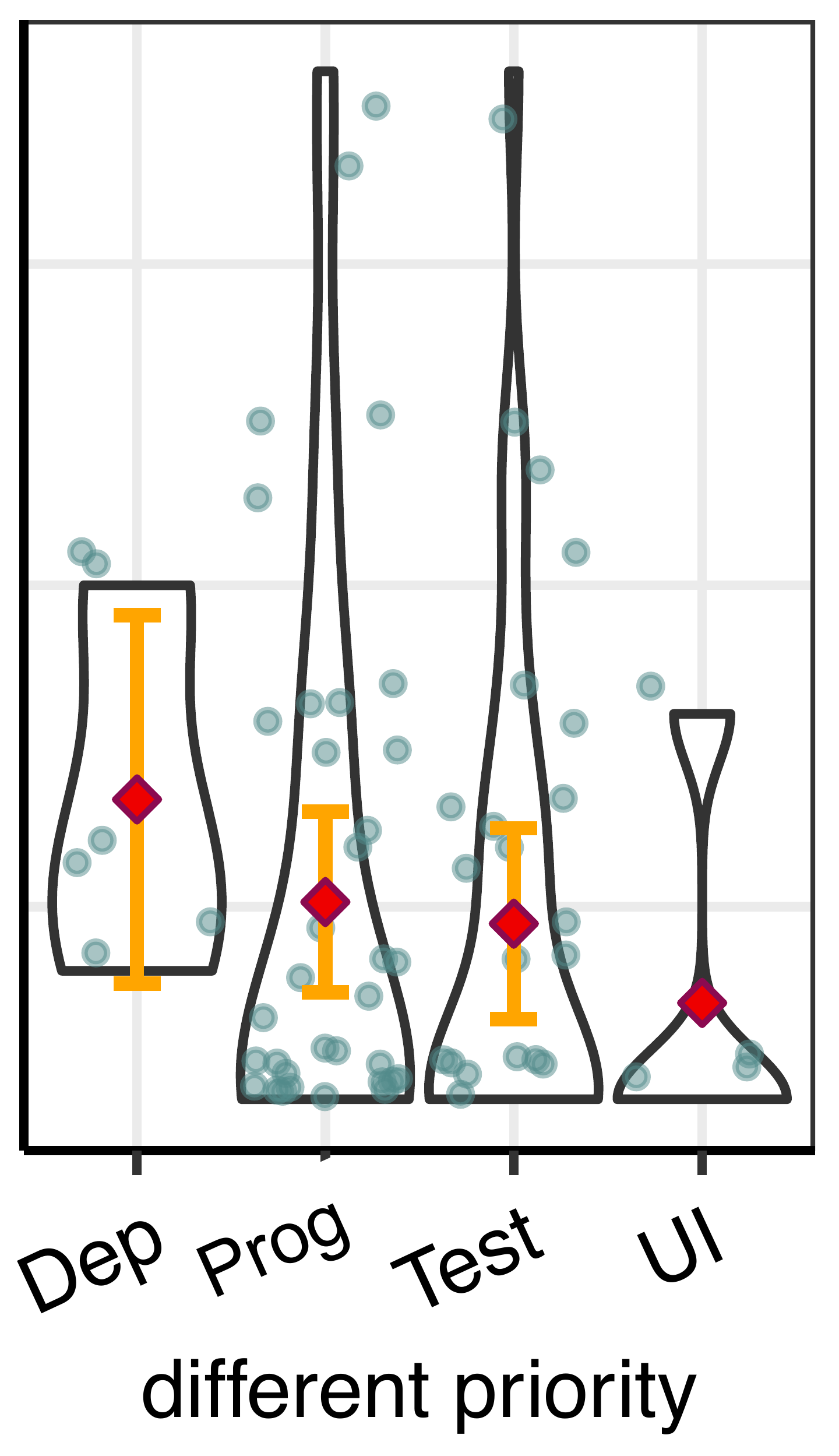}}
\subfloat[]{\includegraphics[scale=0.125]{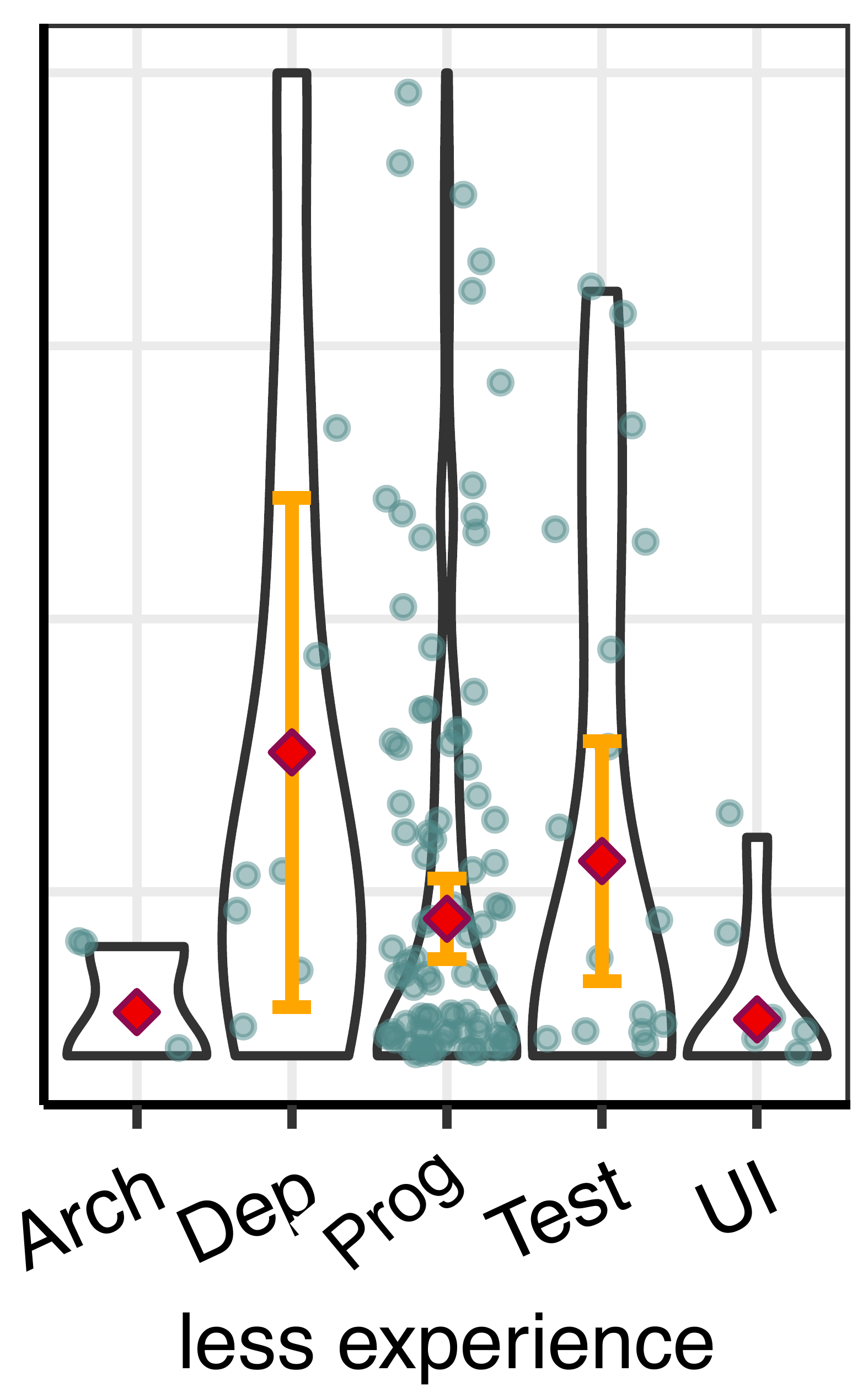}}
\subfloat[]{\includegraphics[scale=0.125]{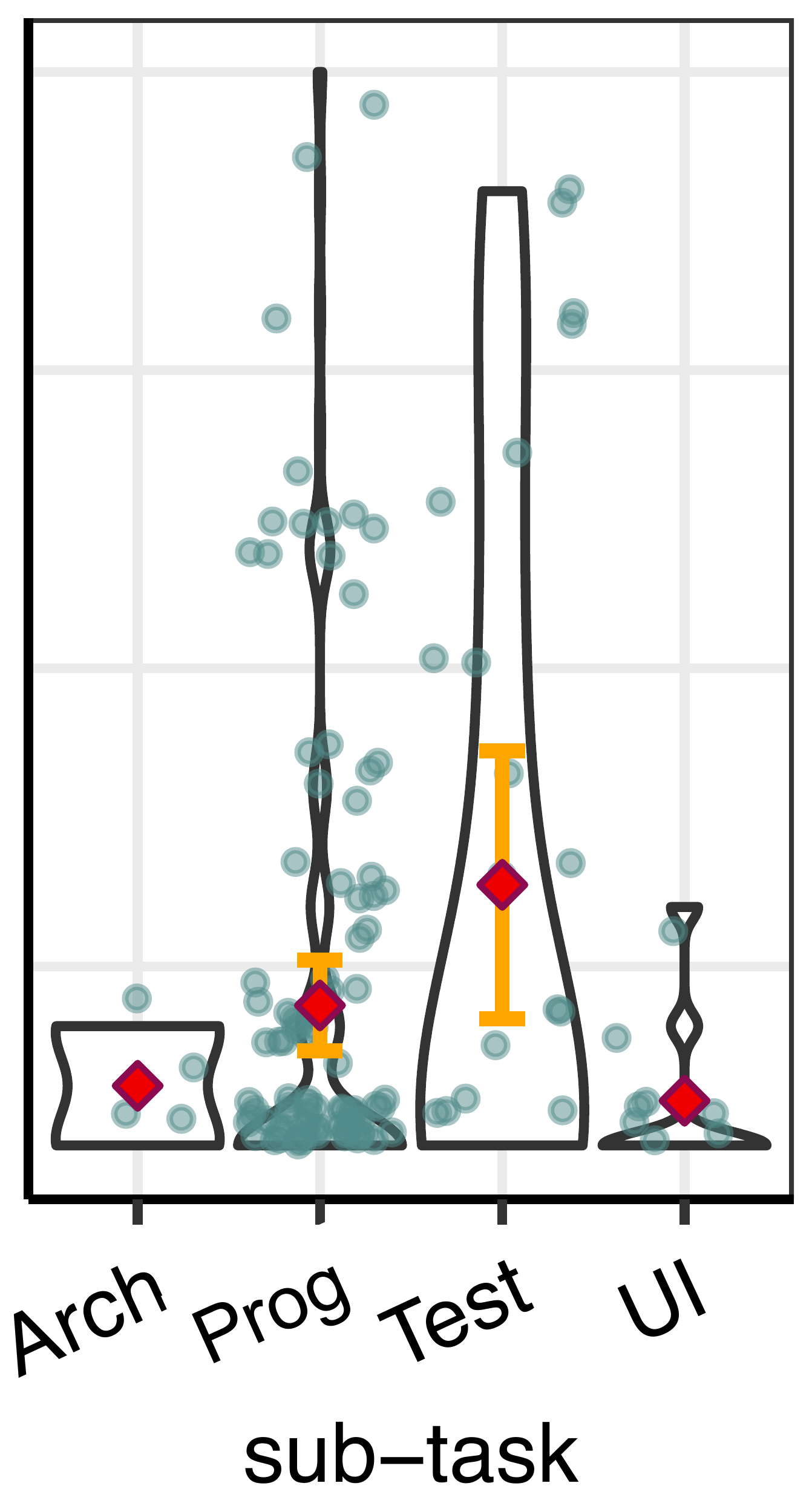}} 
\subfloat[]{\includegraphics[scale=0.125]{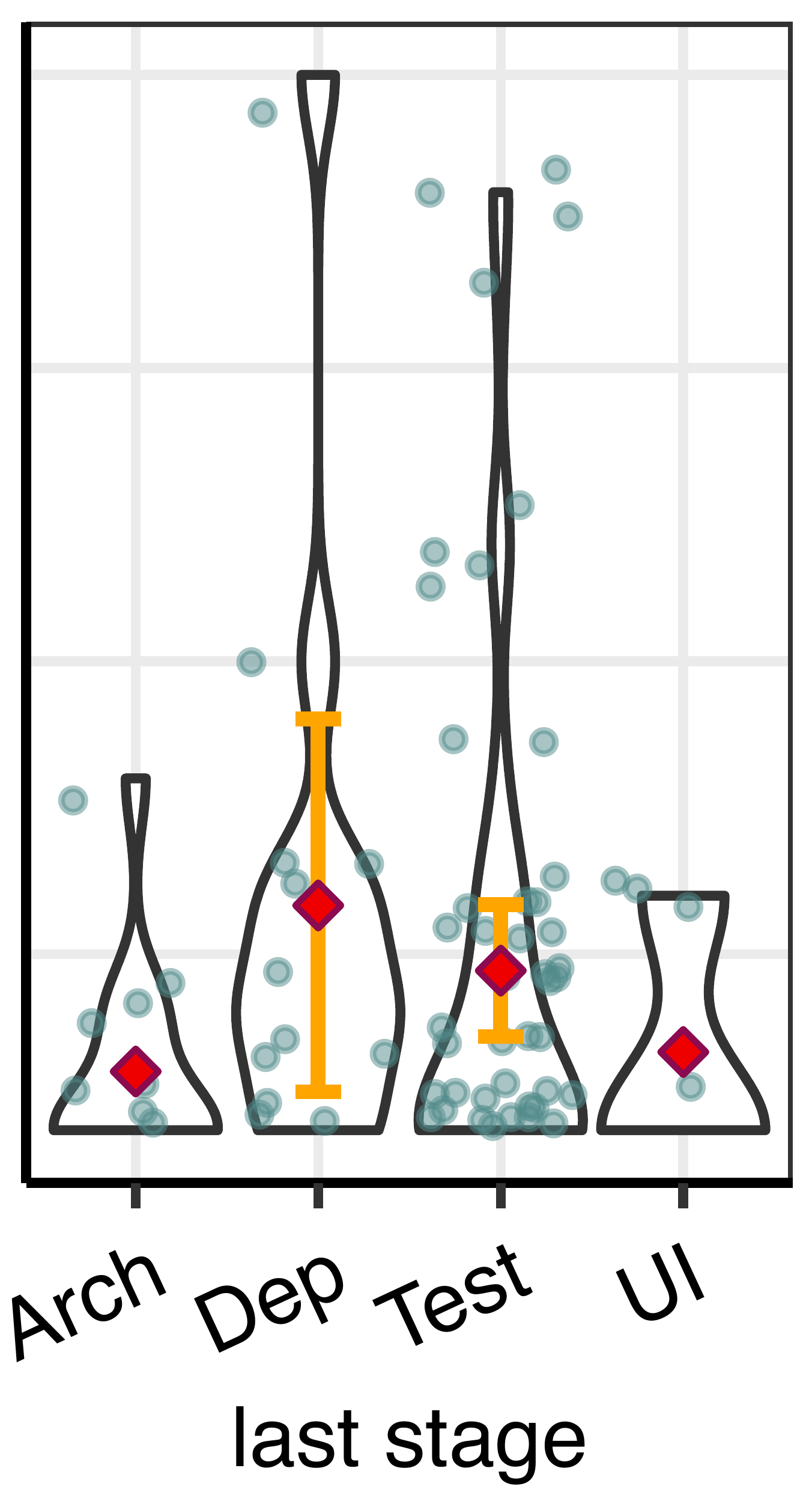}}\\[-2ex]
\subfloat[]{\includegraphics[scale=0.125]{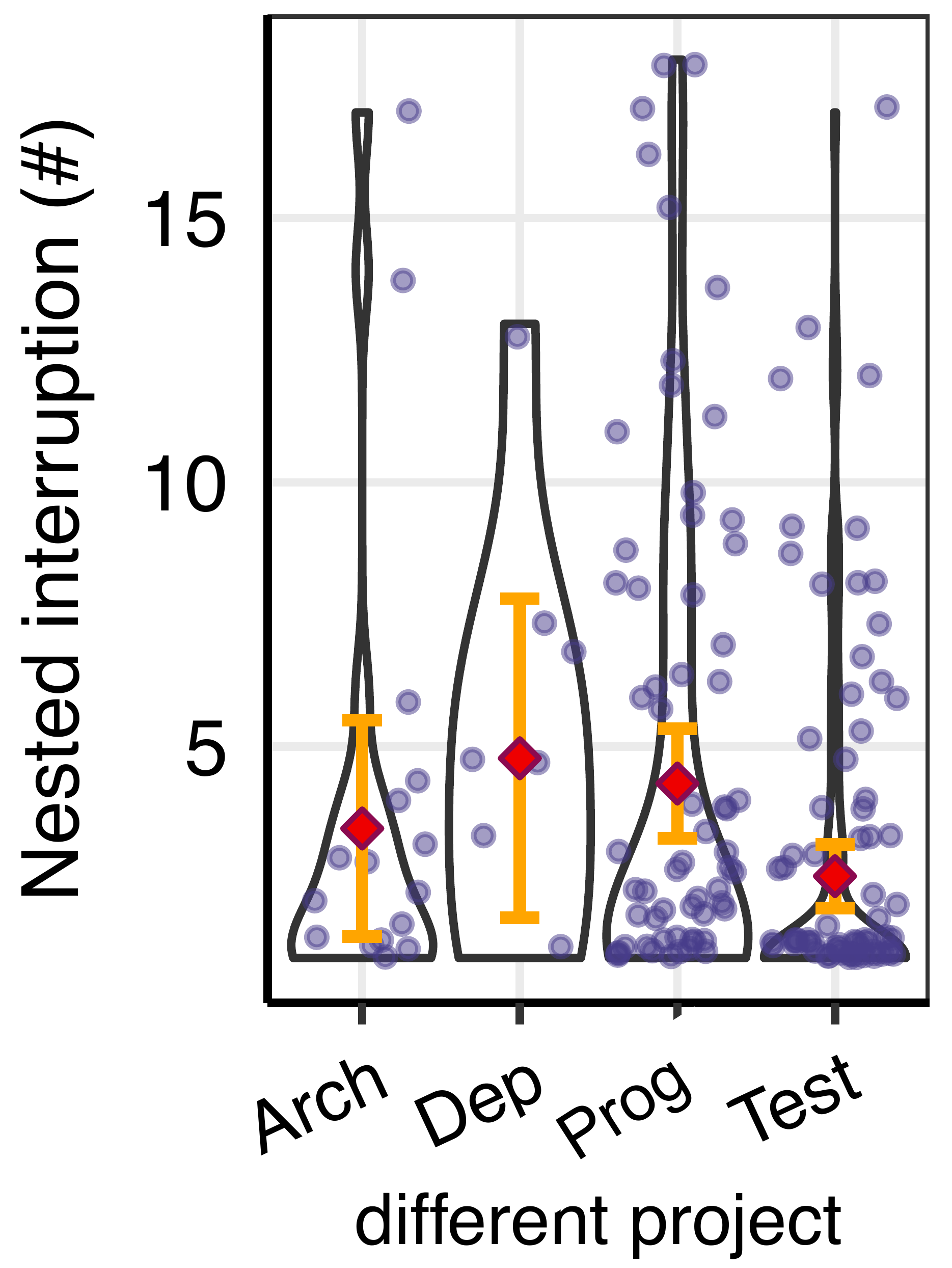}}
\subfloat[]{\includegraphics[scale=0.125]{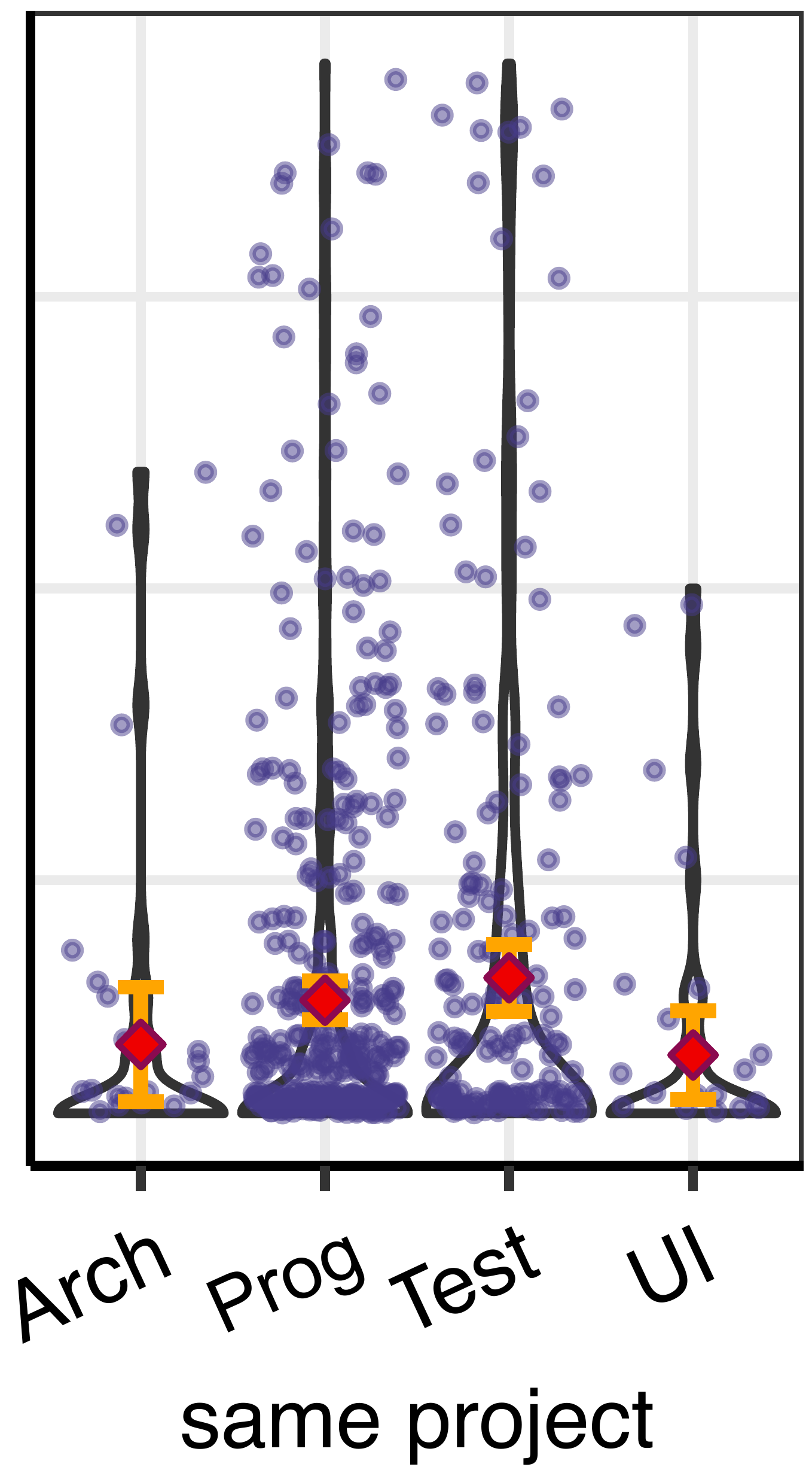}}
\subfloat[]{\includegraphics[scale=0.125]{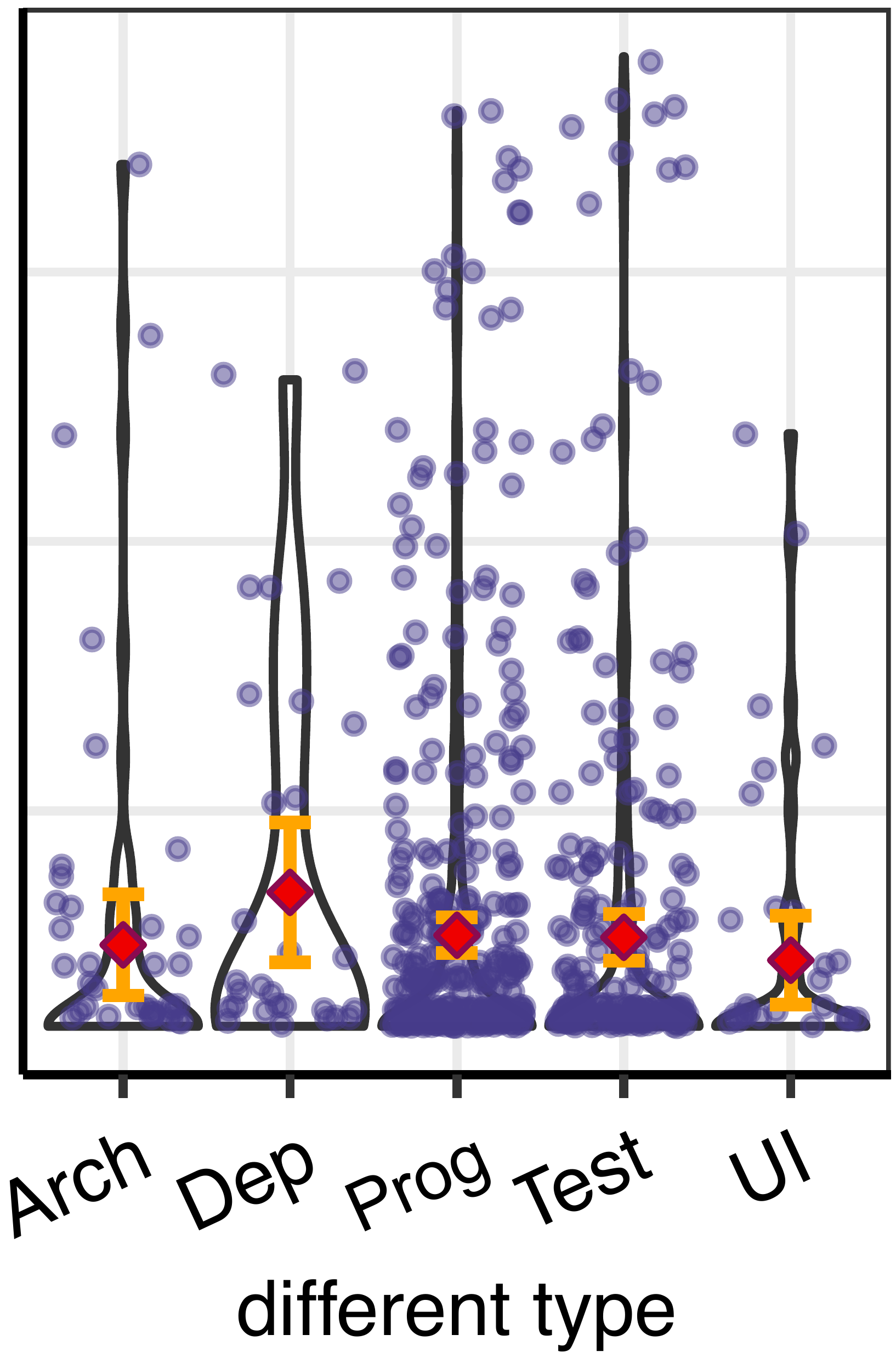}}
\subfloat[]{\includegraphics[scale=0.126]{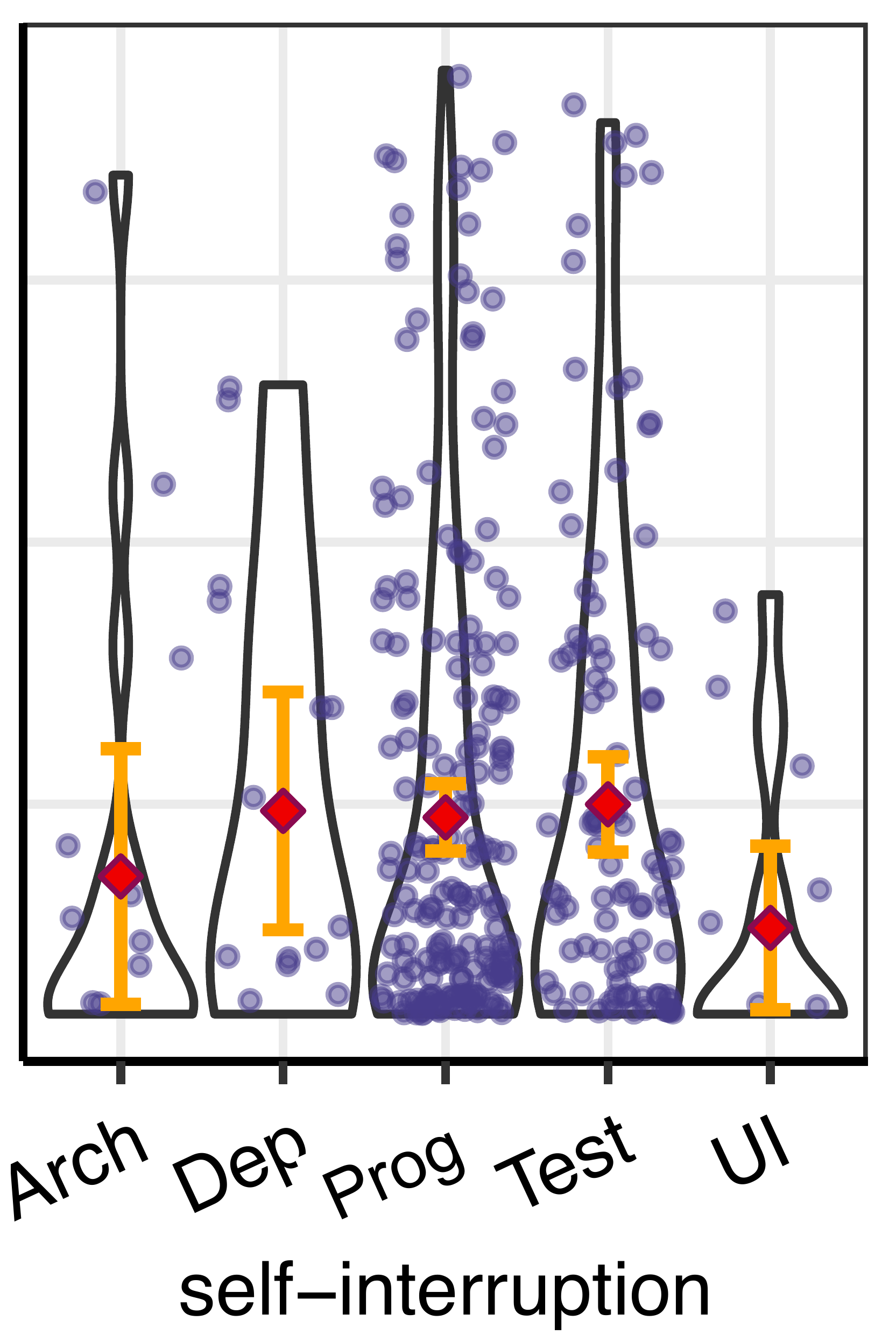}}
\subfloat[]{\includegraphics[scale=0.125]{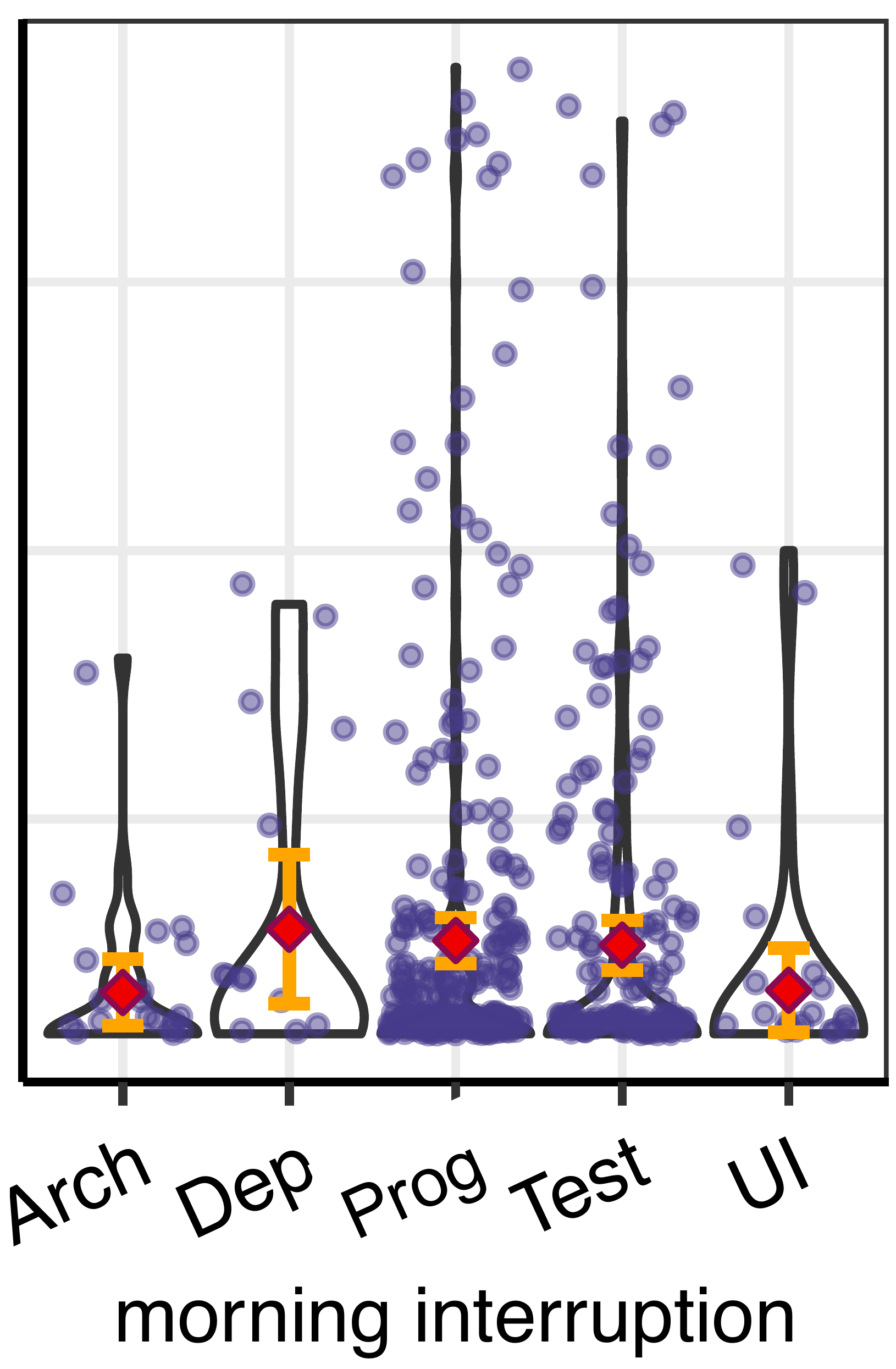}}
\subfloat[]{\includegraphics[scale=0.125]{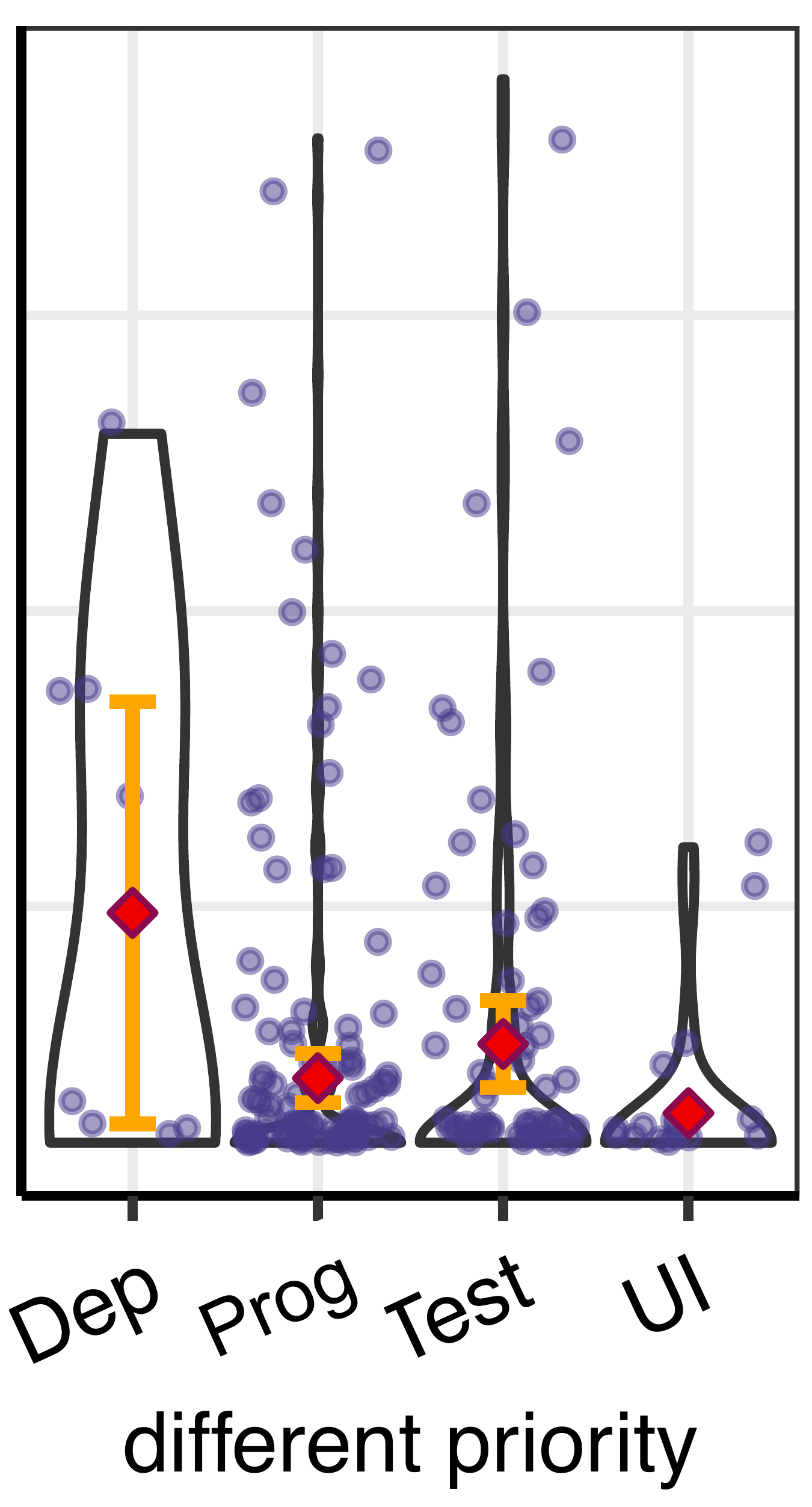}}
\subfloat[]{\includegraphics[scale=0.125]{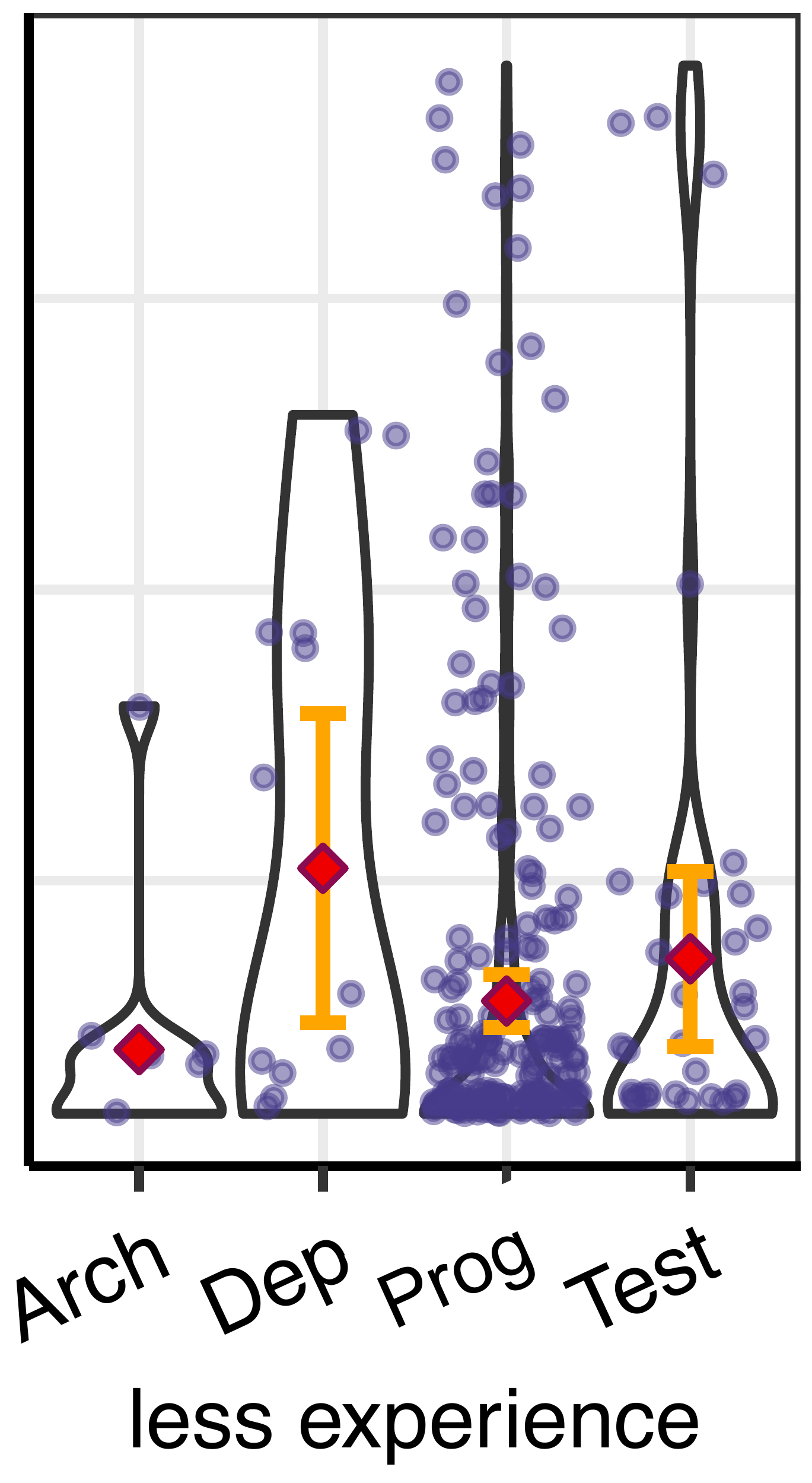}}
\subfloat[]{\includegraphics[scale=0.125]{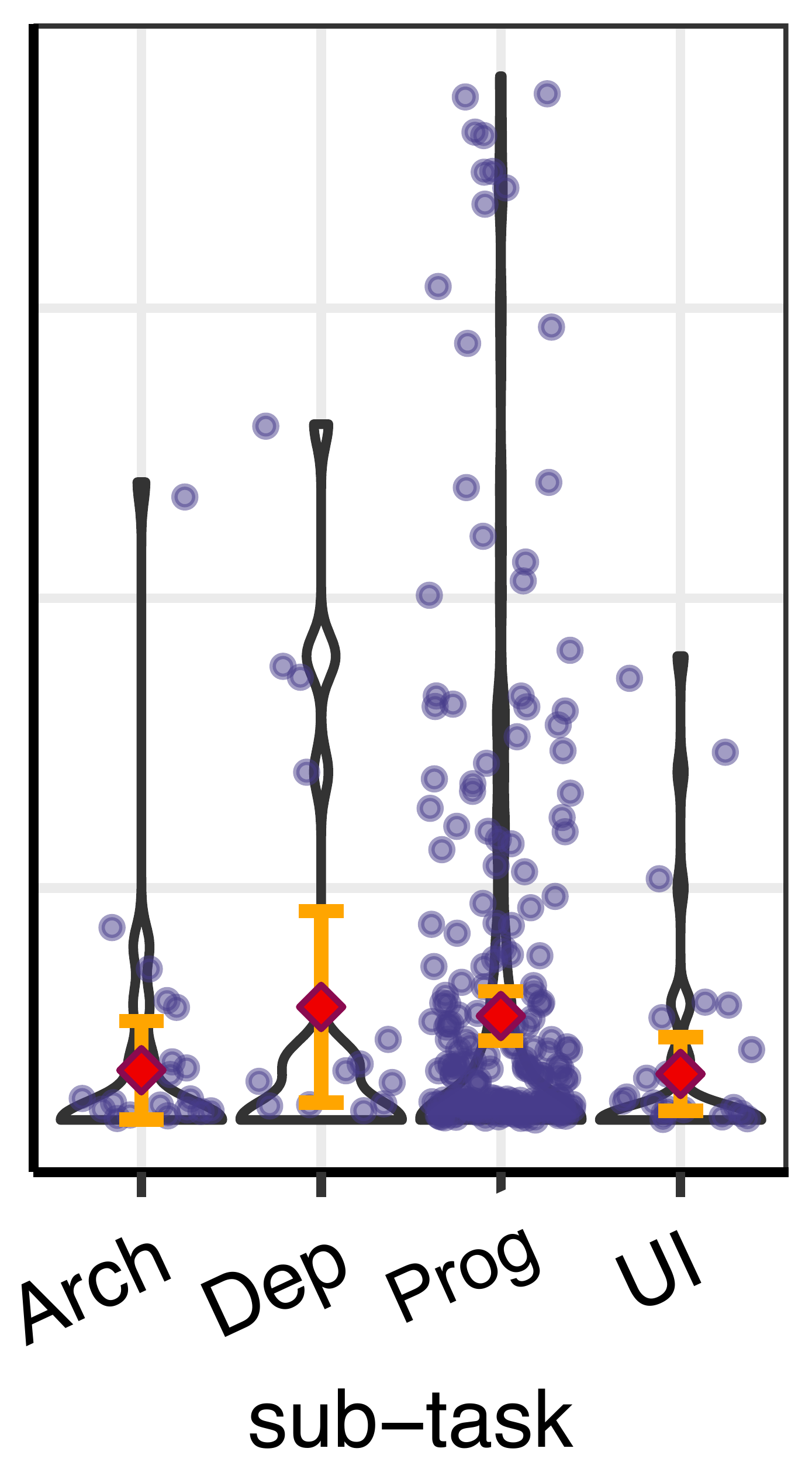}}
\subfloat[]{\includegraphics[scale=0.125]{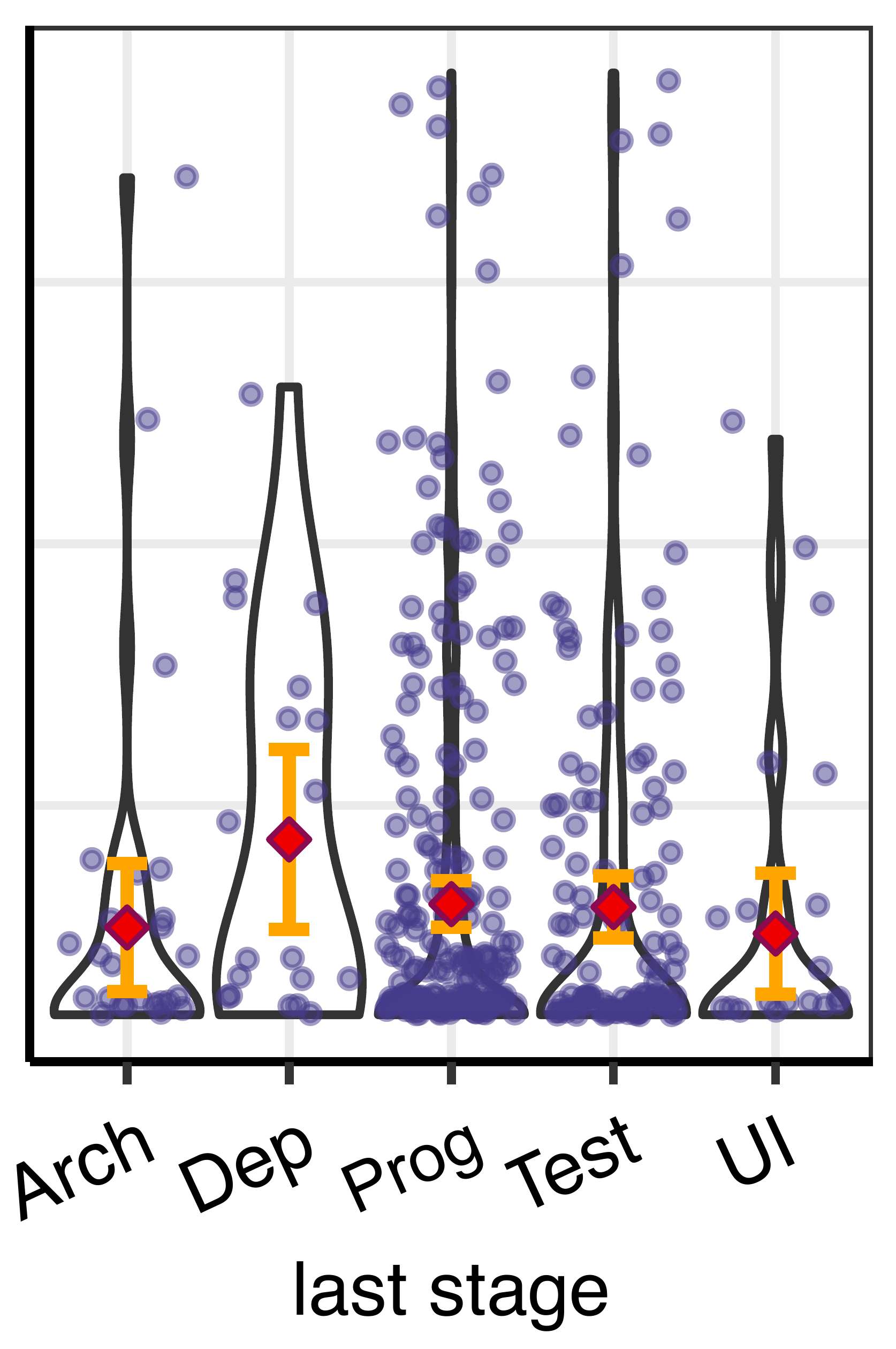}}
\vspace{-3.5mm}
\caption{\small RQ2- 95\% confidence interval of sample means for disruptiveness of interruption characteristics in development tasks }
\label{fig:RQ1}
\vspace{-3mm}
\end{figure*} 
%----------------------------------------------------------
\subsection{RQ2- Comparative Vulnerability}
We posed 160 null hypotheses following this template: 
{\bf \(H_0 =\) The disruptive impact of  \(\imath \nu_i\) {on} \(\Delta\) and/or \(|w|\) { is not different between tasks} \(\langle T, T'\rangle\)}, where \(i \in \{ 1, 2, \dots, 8\}\) and \(\imath\nu_i\), and \(\Delta\)/\(|w|\) denote independent variables and disruptive factors, respectively. \(T\) and \(T'\) represent two different task types for all possible pairs of task types (i.e. \({{5}\choose{2}}=10\) pairs). Table \ref{tab:RQ1} presents the p-value for each of these tests. 
The results of our 95\% confidence interval analysis (e.g. Figure \ref{fig:RQ1}a-q) show that in all cases that task or context-specific factors make a significant difference between deployment and other development tasks, deployment tasks are more vulnerable to interruptions than other task types. 
This could be because deployment tasks are highly interdependent on different tasks within a development process, which makes their resumption more complicated due to the associated tasks.

%--------0-0-0-0-0-0-0-0-0-0-0-0-0-0-0-0-0-0-0-0-0-0-0--------

{\bf Finding \(_{\bf 2-1}\):} The results of Kruskal-Wallis tests show that \emph{priority change} makes a statistically significant difference (\emph{p} =0.002) between the suspension length (\(\Delta\)) for programming and testing tasks (Table \ref{tab:RQ1}). 
Likewise, \emph{experience level} makes a significant difference between the \(\Delta\) and the \(|w|\) of each of {programming and testing} tasks, and UI tasks. 
Regarding the \emph{Task level}, there is a significant difference in \(\Delta\) and \(|w|\) between interrupted low-level programming tasks and each of architecture and UI design tasks. 
There is also a significant difference between switching low-level testing and low-level architectural tasks with respect to suspension length. 
Since the Kruskal-Wallis test only identifies that there is a difference, rather than where the differences lie, we used 95\% confidence intervals (see Figure \ref{fig:RQ1}a-q)  to perform the comparative vulnerability analysis. We use comparison patterns to describe our findings in the following.
 
{\bf Finding \(_{\bf 2-2}\):} For all interruption characteristics (\(\imath\nu_i\)) that make a statistically significant difference between tasks  \(\langle T, T'\rangle\), we provide the following \emph{comparative patterns} for task-specific factors. These patterns compare the vulnerability of two task types \(\langle T, T'\rangle\) to interruption using \(\Delta\) and \(|w|\) measures. 

\begin{framed}
\small
\noindent 
{\bf - Priority Change [PC] ( \(\imath\nu_5\)), (e.g. Figure \ref{fig:RQ1}e) }

 \( \langle PC =1\rangle: \Delta_T>\Delta _{T'} \text{\textcolor{white}{................................}} \text{if }\langle \text{\small Prog}, \text{\small Test}\rangle\)
  \vspace{2.5mm}
  
  \noindent
  {\bf - Experience Level [EL] ( \(\imath\nu_6\)), (e.g. Figure \ref{fig:RQ1}f, o) }

\(  \langle EL =0\rangle:
    \Delta_T>\Delta _{T'}, |w|_T>|w|_{T'}  \text{\textcolor{white}{...}} \text{ if  }\langle \{\text{\small Prog, Test}\}, \text{\small UI}\rangle
\)
  \vspace{2.5mm}
  
  \noindent
  {\bf - Task Level [TL] ( \(\imath\nu_7\)), F(e.g. Figure \ref{fig:RQ1}g, p) }
  
 \(  \langle TL =1\rangle
  \begin{cases}
    \Delta_T>\Delta _{T'}, |w|_T>|w|_{T'}    &\text{if }\langle \text{\small Prog}, \{\text{\small Arch, UI}\}\rangle\\
    \Delta_T>\Delta _{T'} &\text{if }\langle \text{\small Test}, \text{\small Arch}\rangle
  \end{cases}
\)

\end{framed}

\label{sec:D2}

{\bf Finding \(_{\bf 2-3}\):}  We provide the following \emph{comparative patterns} for context-specific factors.
\vspace{-1mm}

\begin{framed}
{\small
\noindent 
{ \bf - Context Switching [CS] ( \(\imath\nu_1\)), (e.g. Figure \ref{fig:RQ1}a-b, i-j) }

 \( \langle CS =1\rangle: \Delta_T>\Delta _{T'}, |w|_T<|w|_{T'}  \text{\textcolor{white}{....}} \text{if }\langle \text{\small Test}, \text{Prog}\rangle\)
   \vspace{2.5mm}
   
  \(  \langle CS =0\rangle
  \begin{cases}
    \Delta_T>\Delta _{T'}, |w|_T>|w|_{T'}    &\text{if }\langle \{\text{\small Prog, Test}\}, \text{\small Arch}\rangle\\
   |w|_T>|w|_{T'} &\text{if }\langle \{\text{\small Prog, Test}\}, \text{\small UI}\rangle
  \end{cases}
\)

  \vspace{3mm}
  
  \noindent
{\bf - Type Difference [TD] ( \(\imath\nu_2\)), F(e.g. Figure \ref{fig:RQ1}c, k) }

  \(  \langle TD =1\rangle
  \begin{cases}
    \Delta_T>\Delta _{T'} & \text{ if  }\langle \text{\small Test}, \text{\small Prog}\rangle\\
    |w|_T>|w|_{T'} &\text { if }\langle \{\text{\small Prog, Test}\}, \text{\small UI}\rangle
  \end{cases}
\)

  \vspace{3mm}

  \noindent
  {\bf - Interruption Type [IT] ( \(\imath\nu_3\)), (e.g. Figure \ref{fig:RQ1}l) }
  
 \(  \langle IT =1\rangle:
     |w|_T>|w|_{T'} \text{\textcolor{white}{........................}}\text{if }\langle \text{\small Prog}, \{\text{\small Arch, UI}\}\rangle
\)\\
{* The  \emph{IT} factor does not make any significant difference of \(\Delta\) between different task types.}
\vspace{3mm}

  \noindent
  { \bf - Daytime [DT] ( \(\imath\nu_4\)), (e.g. Figure \ref{fig:RQ1}d, m)}
  
 \(  \langle DT =1\rangle:
    \Delta_T>\Delta _{T'}, |w|_T>|w|_{T'} \text{\textcolor{white}{.....}}\text{if }\langle \text{\small Prog}, \text{\small UI}\rangle
\)

 \(  \langle DT =0\rangle:
    \Delta_T>\Delta _{T'}\text{\textcolor{white}{..............................}}\text{if }\langle \text{\small Prog}, \text{\small Arch}\rangle
\)
}
\end{framed}
\begin{figure}
\centering
{\includegraphics[scale=0.37]{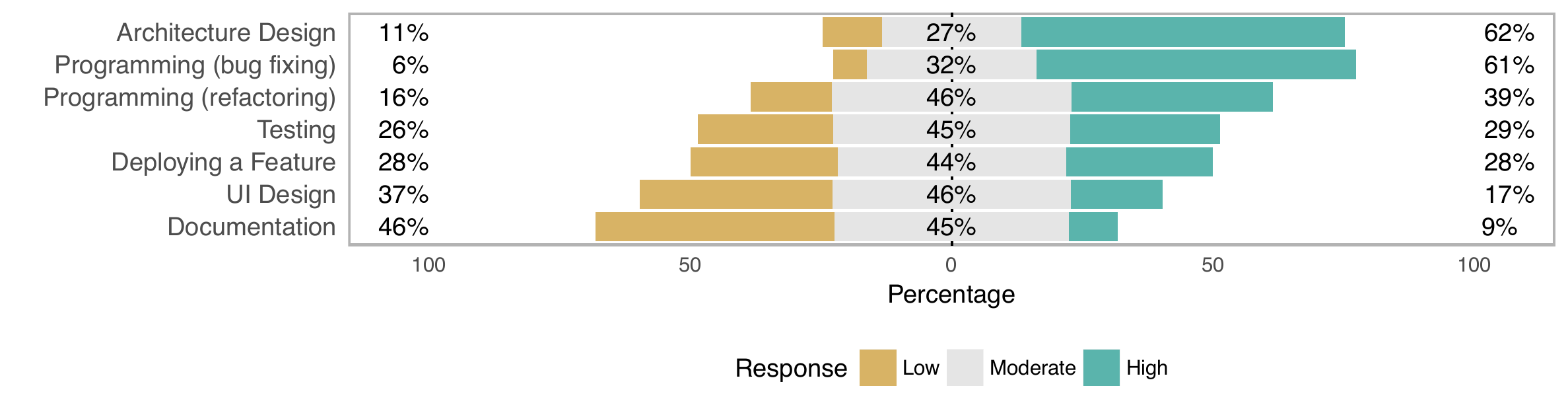}}\\
%\subfloat[Summary of responses grouped by roles]{\includegraphics[scale=0.39]{Figures/E1/LikertRoles}}
\vspace{-3mm}
\caption{Perceived vulnerability of different development tasks to interruption}
\vspace{-5mm}
\label{fig:Likert}
\end{figure}

 {\bf Discussion \(_{\bf 2}\):}
%As illustrated in Table \ref{tab:RQ1}, the \emph{interruption type} does not make any significant difference of \(\Delta\) between different task types. Among six rejected tests for this variables; \(T\in\{\)UI, Arch\(\}\) and  \(T'\in\{\) Dev, Dep, Test\(\}\), looking at Figure \ref{fig:RQ1} (n), task switching makes all \(T'\) tasks are more vulnerable to interruptions by causing greater value for \(|w|\). 
Based on the results of Findings 2-1 and 2-2, in all cases where there is a significant difference between the vulnerability of programming and testing tasks and other task types (\emph{p\(<\)0.05}), these two types are more vulnerable to task switching and interruption. 
This finding is consistent with the experimental evidence and theoretical analysis conducted by Sweller~\cite{PS2}, which shows that solving problems requiring a large number of items be stored in human short-term memory may contribute to excessive cognitive load.
Insofar, as programming and testing tasks require a high number of active statements in developers' \emph{working memory\/}, which contributes to a higher workload, it is reasonable to expect that switching \emph{programming} and \emph{testing} tasks make them more vulnerable to task switching comparing to architectural and UI tasks.  
{However, when we asked survey respondents about the negative impact of task switching/interruption on different types of development tasks (responses are summarized in Figure \ref{fig:Likert}), 117 (89\%) participant reported high or moderate levels of the negative impact of task switching on \emph{architecture design} tasks (i.e. High: 62\%, Moderate: 27\%). \emph{Programming} and \emph{testing} tasks come next, with each of them being 51(\(\pm\)11)\% and 39\% level of agreement.}
However, looking at comparative patterns explored by our retrospective analysis (see Table \ref{tab:RQ1} and Figure \ref{fig:RQ1}), we note that \emph{Architectural} tasks in \emph{all} of the cases are significantly different from other task types and are less vulnerable to interruptions. 
We investigate this difference by conducting a comparison between the survey responses relating to the vulnerability of different development tasks to interruption, grouped by the participants' reported job roles. 
The responses to the task type associated with each job role received higher rating compared to other task types, showing respondent's job role impacts the responses to this question. Moreover, we studied the association between the perceived vulnerability of each task type and the experience level of respondents. 
The results of Spearman's rank correlation tests show that the perceived level of vulnerability ranked by developers does not correlate with their experience level (e.g. Test: rho= 0.13, \(p\)= 0.78).  

Considering the impact of priority change (Finding\(_{2-1}\)), switching to a task with a higher priority makes the suspension period for programming tasks significantly longer than testing tasks (i.e. \(\Delta_{prog}>\Delta_{test}\), \emph{p}=0.002). 
Our survey responses also reflect the perceived negative impact of priority change requests on developers' productivity. 111 (84\%) participants (strongly) agreed with the disruptiveness of unplanned and immediate interruptions such as priority change requests, as in: \emph{``Unplanned requests like high-priority defect fixes don't give me time to save my mental state into the code or the documentation [...] the less likely I can return easily''}. Conversely, compared to programming tasks, testing tasks are more vulnerable to context and type switching (Figure \ref{fig:RQ1}a-c), as stated by one of our survey participants: \emph{``As testing can take a different type of mindset than a typical development phase, if switching occurs at mid-task collecting thoughts to return to the task's context can be disruptive and time-consuming''}.

%Taking all retrospective and survey data and existing evidence together, d
{\bf Practitioner's corner \(_{\bf 2}\):} Due to the problem-solving nature of programming and testing tasks, and knowing that human short-term memory is severely limited \cite{PS2, MofGoals} and cannot accommodate a large number of items, we recommend practitioners minimize switching programming and testing tasks. 
Further, considering that testing tasks are more vulnerable to context-switching than programming, architecture, and UI design tasks, we propose that it might be more efficient if testers minimize their project switches or they respond to fewer context-switching requests.

 %----------------------------------------------------------

%=================================================================================
 
 \begin{figure}
\includegraphics[scale=0.42]{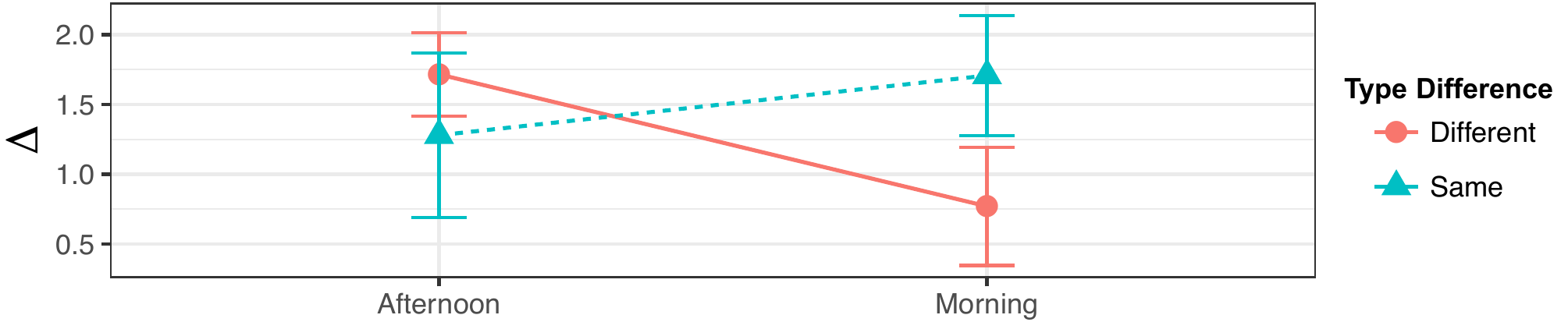}
\vspace{-5mm}
\caption{\small 95\% confidence interval of TD/DT factors interaction (Scheirer-Ray-Hare test: \(p\)=0.01)}
\label{fig:Interaction}
\vspace{-5mm}
\end{figure}

\begin {table*}
\scriptsize
\centering
\caption {\small RQ3- Two-way factorial Scheirer-Ray-Hare Test. [CS]: Context Switching, [TD]: Type Difference, [IT]: Interruption Type, [DT]: Daytime, [PC]: Priority Change, [EL]: Experience Level, [TL]: Task Level, [TS]: Task Stage. }
\vspace{-2.5mm}
\label{tab:RQ3}
\setlength{\tabcolsep}{.15em}
\begin{tabular} {|p{0.8cm}|p{0.75cm}!{\color{black}\vrule}p{.25cm}p{0.25cm}|p{0.75cm}!{\color{black}\vrule}p{1.2cm}p{1.2cm}|p{.75cm}!{\color{black}\vrule}p{1.25cm}p{1.2cm}|p{0.75cm}!{\color{black}\vrule}p{1cm}p{.25cm}|p{0.75cm}!{\color{black}\vrule}p{0.25cm}p{.25cm}p{.25cm}|} \hline
&\multicolumn{3}{c|}{{\bf Architecture}}&\multicolumn{3}{c|}{\bf Programming}&\multicolumn{3}{c|}{\bf Test}&\multicolumn{3}{c|}{{\bf UI}}&\multicolumn{3}{c|}{{\bf Deployment}}\\\cline{2-16}
\multicolumn{1}{|c|}{\bf Pairs}&\multicolumn{1}{c|}{\(\phi\)}&\multicolumn{1}{c}{\(\Delta\)}&\multicolumn{1}{c|}{\(|w|\)
}&\multicolumn{1}{c|}{\(\phi\)}& \multicolumn{1}{c}{\(\Delta\)}&\multicolumn{1}{c|}{\(|w|\)
}&\multicolumn{1}{c|}{\(\phi\)}& \multicolumn{1}{c}{\(\Delta\)}&\multicolumn{1}{c|}{\(|w|\)
}& \multicolumn{1}{c|}{\(\phi\)}& \multicolumn{1}{c}{\(\Delta\)}&\multicolumn{1}{c|}{\(|w|\)
}&\multicolumn{1}{c|}{\(\phi\)}& \multicolumn{1}{c}{\(\Delta\)}&\multicolumn{1}{c|}{\(|w|\)
}\\\hline
\hline
\cellcolor{g1}{\bf CS-EL}&\textcolor{white}{-}0.42&\multicolumn{1}{c}{\ding{121}}	&\multicolumn{1}{c|}{\ding{121}}&\cellcolor{gray}\textcolor{gray}{-}0.79\textcolor{gray}{..}\textcolor{blue2}{\ding{108}}&\multicolumn{2}{c|}{\textcolor{g1}{\ding{121} \ding{121} \ding{121} \ding{121} \ding{121} \ding{121} \ding{121} \ding{121} \ding{121} \ding{121} \ding{121} \ding{121} \ding{121} \ding{121} \ding{121} \ding{121} \ding{121} \ding{121} \ding{121} \ding{121} \ding{121} \ding{121} \ding{121}}}&0.001&\multicolumn{2}{c|}{\textcolor{g1}{\ding{121} \ding{121} \ding{121} \ding{121} \ding{121} \ding{121} \ding{121} \ding{121} \ding{121} \ding{121} \ding{121} \ding{121} \ding{121} \ding{121} \ding{121} \ding{121} \ding{121} \ding{121} \ding{121} \ding{121} \ding{121} \ding{121} \ding{121}}}&\cellcolor{gray}-0.90\textcolor{gray}{..}\textcolor{red3}{  \ding{108}}&\multicolumn{1}{c}{\textcolor{g1}{\ding{121} \ding{121} \ding{121} \ding{121} \ding{121} \ding{121} \ding{121} \ding{121} \ding{121} \ding{121} \ding{121}}}&\multicolumn{1}{c|}{\ding{121}}&\cellcolor{gray}\textcolor{gray}{-}0.80\textcolor{gray}{..}\textcolor{blue3}{\ding{108}}&\multicolumn{1}{c}{\ding{121}}&\multicolumn{1}{c|}{\ding{121}}\\

\cellcolor{g1}{\bf CS-TL}&\cellcolor{gray}-0.64\textcolor{gray}{..}\textcolor{red1}{  \ding{108}}&\multicolumn{1}{c}{\ding{121}}&\multicolumn{1}{c|}{\ding{121}}&\cellcolor{gray}-0.80\textcolor{gray}{..}\textcolor{red3}{  \ding{108}}&\multicolumn{2}{c|}{\textcolor{g1}{\ding{121} \ding{121} \ding{121} \ding{121} \ding{121} \ding{121} \ding{121} \ding{121} \ding{121} \ding{121} \ding{121} \ding{121} \ding{121} \ding{121} \ding{121} \ding{121} \ding{121} \ding{121} \ding{121} \ding{121} \ding{121} \ding{121} \ding{121}}}&-0.13&\multicolumn{2}{c|}{\textcolor{g1}{\ding{121} \ding{121} \ding{121} \ding{121} \ding{121} \ding{121} \ding{121} \ding{121} \ding{121} \ding{121} \ding{121} \ding{121} \ding{121} \ding{121} \ding{121} \ding{121} \ding{121} \ding{121} \ding{121} \ding{121} \ding{121} \ding{121} \ding{121}}}&\cellcolor{gray} -0.91\textcolor{gray}{..}\textcolor{red3}{  \ding{108}}&\multicolumn{1}{c}{\textcolor{g1}{\ding{121} \ding{121} \ding{121} \ding{121} \ding{121} \ding{121} \ding{121} \ding{121} \ding{121} \ding{121} \ding{121}}}&\multicolumn{1}{c|}{\ding{121}}&\cellcolor{gray}-0.90\textcolor{gray}{..}\textcolor{red3}{\ding{108}}&\multicolumn{1}{c}{\ding{121}}&\multicolumn{1}{c|}{\ding{121}}\\
{\bf CS-TD}&\textcolor{white}{-}0.11&\multicolumn{1}{c}{\ding{121}}&\multicolumn{1}{c|}{\ding{121}}&\textcolor{white}{-}0.02&&&-0.33&\cellcolor{gray}1e-3 \(\langle 10, 01 \rangle\)&&\cellcolor{gray}\textcolor{gray}{-}0.62\textcolor{gray}{..}\textcolor{blue1}{  \ding{108}}&&\multicolumn{1}{c|}{\ding{121}}&\cellcolor{gray} \textcolor{gray}{-}0.84\textcolor{gray}{..}\textcolor{blue3}{  \ding{108}}&\multicolumn{1}{c}{\ding{121}}&\multicolumn{1}{c|}{\ding{121}}\\

{\bf CS-IT}&\textcolor{white}{-}0.05&\multicolumn{1}{c}{\ding{121}}&\multicolumn{1}{c|}{\ding{121}}&\textcolor{white}{-}0.10&&&\textcolor{white}{-}0.21&\cellcolor{gray}0.02 \(\langle 10, 01 \rangle\)&&-0.10&&\multicolumn{1}{c|}{\ding{121}}&\cellcolor{gray}\textcolor{gray}{-}0.56\textcolor{gray}{..}\textcolor{blue1}{  \ding{108}}&\multicolumn{1}{c}{\ding{121}}&\multicolumn{1}{c|}{\ding{121}}\\
{\bf CS-PC}&\cellcolor{gray}-0.76\textcolor{gray}{..}\textcolor{red2}{  \ding{108}}&\multicolumn{1}{c}{\ding{121}}&\multicolumn{1}{c|}{\ding{121}}&\textcolor{white}{-}0.36&\multicolumn{1}{c}{\cellcolor{gray}0.04 \(\langle10,01\rangle\)}&&-0.46&&\multicolumn{1}{c|}{\cellcolor{gray}0.02 \(\langle 10, 01 \rangle\)}&\textcolor{white}{-}0.24&&\multicolumn{1}{c|}{\ding{121}}&\textcolor{white}{-}0.48&\multicolumn{1}{c}{\ding{121}}&\multicolumn{1}{c|}{\ding{121}}\\

\cellcolor{g1}{\bf CS-DT}&\cellcolor{gray}-0.72\textcolor{gray}{..}\textcolor{red2}{  \ding{108}}&\multicolumn{1}{c}{\ding{121}}&\multicolumn{1}{c|}{\ding{121}}&-0.10&\multicolumn{2}{c|}{\textcolor{g1}{\ding{121} \ding{121} \ding{121} \ding{121} \ding{121} \ding{121} \ding{121} \ding{121} \ding{121} \ding{121} \ding{121} \ding{121} \ding{121} \ding{121} \ding{121} \ding{121} \ding{121} \ding{121} \ding{121} \ding{121} \ding{121} \ding{121} \ding{121}}}&\textcolor{white}{-}0.14&\multicolumn{2}{c|}{\textcolor{g1}{\ding{121} \ding{121} \ding{121} \ding{121} \ding{121} \ding{121} \ding{121} \ding{121} \ding{121} \ding{121} \ding{121} \ding{121} \ding{121} \ding{121} \ding{121} \ding{121} \ding{121} \ding{121} \ding{121} \ding{121} \ding{121} \ding{121} \ding{121}}}&-0.49&\multicolumn{1}{c}{\textcolor{g1}{\ding{121} \ding{121} \ding{121} \ding{121} \ding{121} \ding{121} \ding{121} \ding{121} \ding{121} \ding{121} \ding{121}}}&\multicolumn{1}{c|}{\ding{121}}&-0.31&\multicolumn{1}{c}{\ding{121}}&\multicolumn{1}{c|}{\ding{121}}\\

\cellcolor{g1}{\bf CS-TS}&-0.42&\multicolumn{1}{c}{\ding{121}}&\multicolumn{1}{c|}{\ding{121}}&-0.44&\multicolumn{2}{c|}{\textcolor{g1}{\ding{121} \ding{121} \ding{121} \ding{121} \ding{121} \ding{121} \ding{121} \ding{121} \ding{121} \ding{121} \ding{121} \ding{121} \ding{121} \ding{121} \ding{121} \ding{121} \ding{121} \ding{121} \ding{121} \ding{121} \ding{121} \ding{121} \ding{121}}}&\textcolor{white}{-}0.36&\multicolumn{2}{c|}{\textcolor{g1}{\ding{121} \ding{121} \ding{121} \ding{121} \ding{121} \ding{121} \ding{121} \ding{121} \ding{121} \ding{121} \ding{121} \ding{121} \ding{121} \ding{121} \ding{121} \ding{121} \ding{121} \ding{121} \ding{121} \ding{121} \ding{121} \ding{121} \ding{121}}}&\textcolor{white}{-}0.12&\multicolumn{1}{c}{\textcolor{g1}{\ding{121} \ding{121} \ding{121} \ding{121} \ding{121} \ding{121} \ding{121} \ding{121} \ding{121} \ding{121} \ding{121} }}&\multicolumn{1}{c|}{\ding{121}}&\cellcolor{gray}-0.83&\multicolumn{1}{c}{\ding{121}}&\multicolumn{1}{c|}{\ding{121}}\\\hline\hline

{\bf EL-TL}&\cellcolor{gray}-0.80\textcolor{gray}{..}\textcolor{red3}{  \ding{108}}&\multicolumn{1}{c}{\ding{121}}&\multicolumn{1}{c|}{\ding{121}}&\cellcolor{gray}-0.98\textcolor{gray}{..}\textcolor{red3}{  \ding{108}}&&&-0.10&&&\cellcolor{gray}-0.99\textcolor{gray}{..}\textcolor{red3}{  \ding{108}}&&\multicolumn{1}{c|}{\ding{121}}&\cellcolor{gray}-0.84\textcolor{gray}{..}\textcolor{red3}{  \ding{108}}&\multicolumn{1}{c}{\ding{121}}&\multicolumn{1}{c|}{\ding{121}}\\

{\bf EL-TD}&\cellcolor{gray}-0.78\textcolor{gray}{..}\textcolor{red2}{  \ding{108}}&\multicolumn{1}{c}{\ding{121}}&\multicolumn{1}{c|}{\ding{121}}&-0.47&\multicolumn{1}{c}{\cellcolor{gray}0.00 \(\langle 00, 11 \rangle\)}&&-0.38&&&\textcolor{white}{-}0.28&\multicolumn{1}{c}{\cellcolor{gray}0.03 \(\langle 11, 00 \rangle\)}&\multicolumn{1}{c|}{\ding{121}}&\cellcolor{gray}\textcolor{gray}{-}0.59\textcolor{gray}{..}\textcolor{blue1}{  \ding{108}}&\multicolumn{1}{c}{\ding{121}}&\multicolumn{1}{c|}{\ding{121}}\\
{\bf EL-IT}&-0.48&\multicolumn{1}{c}{\ding{121}}&\multicolumn{1}{c|}{\ding{121}}&\cellcolor{gray}\textcolor{gray}{-}0.63\textcolor{gray}{..}\textcolor{blue1}{  \ding{108}}&\multicolumn{1}{c}{\cellcolor{gray}0.01 \(\langle 01, 10 \rangle\)}&&\cellcolor{gray}-0.53\textcolor{gray}{..}\textcolor{red1}{  \ding{108}}&&\multicolumn{1}{c|}{\cellcolor{gray}0.01 \(\langle 00, 11 \rangle\)}&-0.39&&\multicolumn{1}{c|}{\ding{121}}&\textcolor{white}{-}0.36&\multicolumn{1}{c}{\ding{121}}&\multicolumn{1}{c|}{\ding{121}}\\

{\bf EL-PC}&-0.30&\multicolumn{1}{c}{\ding{121}}&\multicolumn{1}{c|}{\ding{121}}&\cellcolor{gray}\textcolor{gray}{-}0.77\textcolor{gray}{..}\textcolor{blue2}{  \ding{108}}&\multicolumn{1}{c}{\cellcolor{gray}0.02 \(\langle 01, 10 \rangle\)}&\multicolumn{1}{c|}{\cellcolor{gray}0.01 \(\langle 01, 10 \rangle\)
}&-0.11&&&\cellcolor{gray}\textcolor{gray}{-}0.59\textcolor{gray}{..}\textcolor{blue1}{  \ding{108}}&&\multicolumn{1}{c|}{\ding{121}}&\textcolor{white}{-}0.25&\multicolumn{1}{c}{\ding{121}}&\multicolumn{1}{c|}{\ding{121}}\\
\cellcolor{g1}{\bf EL-DT}&\cellcolor{gray}-0.59\textcolor{gray}{..}\textcolor{red1}{  \ding{108}}&\multicolumn{1}{c}{\ding{121}}&\multicolumn{1}{c|}{\ding{121}}&-0.27&\multicolumn{2}{c|}{\textcolor{g1}{\ding{121} \ding{121} \ding{121} \ding{121} \ding{121} \ding{121} \ding{121} \ding{121} \ding{121} \ding{121} \ding{121} \ding{121} \ding{121} \ding{121} \ding{121} \ding{121} \ding{121} \ding{121} \ding{121} \ding{121} \ding{121} \ding{121} \ding{121}}}&\cellcolor{gray}-0.70\textcolor{gray}{..}\textcolor{red2}{  \ding{108}}&\multicolumn{2}{c|}{\textcolor{g1}{\ding{121} \ding{121} \ding{121} \ding{121} \ding{121} \ding{121} \ding{121} \ding{121} \ding{121} \ding{121} \ding{121} \ding{121} \ding{121} \ding{121} \ding{121} \ding{121} \ding{121} \ding{121} \ding{121} \ding{121} \ding{121} \ding{121} \ding{121}}}&\cellcolor{gray}-0.63\textcolor{gray}{..}\textcolor{red1}{  \ding{108}}&\multicolumn{1}{c}{\textcolor{g1}{\ding{121} \ding{121} \ding{121} \ding{121} \ding{121} \ding{121} \ding{121} \ding{121} \ding{121} \ding{121} \ding{121}}}&\multicolumn{1}{c|}{\ding{121}}&-0.42&\multicolumn{1}{c}{\ding{121}}&\multicolumn{1}{c|}{\ding{121}}\\

\cellcolor{g1}{\bf EL-TS}&\cellcolor{gray}-0.50\textcolor{gray}{..}\textcolor{red1}{  \ding{108}}&\multicolumn{1}{c}{\ding{121}}&\multicolumn{1}{c|}{\ding{121}}&-0.25&\multicolumn{2}{c|}{\textcolor{g1}{\ding{121} \ding{121} \ding{121} \ding{121} \ding{121} \ding{121} \ding{121} \ding{121} \ding{121} \ding{121} \ding{121} \ding{121} \ding{121} \ding{121} \ding{121} \ding{121} \ding{121} \ding{121} \ding{121} \ding{121} \ding{121} \ding{121} \ding{121}}}&\cellcolor{gray}\textcolor{gray}{-}0.62\textcolor{gray}{..}\textcolor{blue1}{  \ding{108}}&\multicolumn{2}{c|}{\textcolor{g1}{\ding{121} \ding{121} \ding{121} \ding{121} \ding{121} \ding{121} \ding{121} \ding{121} \ding{121} \ding{121} \ding{121} \ding{121} \ding{121} \ding{121} \ding{121} \ding{121} \ding{121} \ding{121} \ding{121} \ding{121} \ding{121} \ding{121} \ding{121}}}&\textcolor{white}{-}0.36&\multicolumn{1}{c}{\textcolor{g1}{\ding{121} \ding{121} \ding{121} \ding{121} \ding{121} \ding{121} \ding{121} \ding{121} \ding{121} \ding{121} \ding{121}}}&\multicolumn{1}{c|}{\ding{121}}&\cellcolor{gray}-0.78\textcolor{gray}{..}\textcolor{red2}{  \ding{108}}&\multicolumn{1}{c}{\ding{121}}&\multicolumn{1}{c|}{\ding{121}}\\\hline\hline

{\bf TL-TD}&\textcolor{white}{-}0.47&\multicolumn{1}{c}{\ding{121}}&\multicolumn{1}{c|}{\ding{121}}&\textcolor{white}{-}0.39&\multicolumn{1}{c}{\cellcolor{gray}4e-4 \(\langle 01, 10 \rangle\)}&\multicolumn{1}{c|}{\cellcolor{gray}0.01 \(\langle 01, 10 \rangle\)}&-0.23&&&-0.28&\multicolumn{1}{c}{\cellcolor{gray}0.02 \(\langle 01, 10 \rangle\)}&\multicolumn{1}{c|}{\ding{121}}&\cellcolor{gray}-0.63\textcolor{gray}{..}\textcolor{red1}{  \ding{108}}&\multicolumn{1}{c}{\ding{121}}&\multicolumn{1}{c|}{\ding{121}}\\

{\bf TL-IT}&\cellcolor{gray}\textcolor{gray}{-}0.55\textcolor{gray}{..}\textcolor{blue1}{  \ding{108}}&\multicolumn{1}{c}{\ding{121}}&\multicolumn{1}{c|}{\ding{121}}&\textcolor{white}{-}0.23&\multicolumn{1}{c}{\cellcolor{gray}0.003 \(\langle 00, 11 \rangle\)}&&-0.22&&&\textcolor{white}{-}0.40&\multicolumn{1}{c}{\cellcolor{gray}0.04 \(\langle 00, 11 \rangle\)}&\multicolumn{1}{c|}{\ding{121}}&-0.43&\multicolumn{1}{c}{\ding{121}}&\multicolumn{1}{c|}{\ding{121}}\\

{\bf TL-PC}&\textcolor{white}{-}0.28&\multicolumn{1}{c}{\ding{121}}&\multicolumn{1}{c|}{\ding{121}}&\cellcolor{gray}-0.73\textcolor{gray}{..}\textcolor{red2}{  \ding{108}}&&\multicolumn{1}{c|}{\cellcolor{gray}0.03 \(\langle 00, 11 \rangle\)}&-0.29&&&\cellcolor{gray}-0.60\textcolor{gray}{..}\textcolor{red2}{  \ding{108}}&&\multicolumn{1}{c|}{\ding{121}}&-0.39&\multicolumn{1}{c}{\ding{121}}&\multicolumn{1}{c|}{\ding{121}}\\

\cellcolor{g1}{\bf TL-DT}&\cellcolor{gray}\textcolor{gray}{-}0.51\textcolor{gray}{..}\textcolor{blue1}{  \ding{108}}&\multicolumn{1}{c}{\ding{121}}&\multicolumn{1}{c|}{\ding{121}}&\textcolor{white}{-}0.15&\multicolumn{2}{c|}{\textcolor{g1}{\ding{121} \ding{121} \ding{121} \ding{121} \ding{121} \ding{121} \ding{121} \ding{121} \ding{121} \ding{121} \ding{121} \ding{121} \ding{121} \ding{121} \ding{121} \ding{121} \ding{121} \ding{121} \ding{121} \ding{121} \ding{121} \ding{121} \ding{121}}}&-0.10&\multicolumn{2}{c|}{\textcolor{g1}{\ding{121} \ding{121} \ding{121} \ding{121} \ding{121} \ding{121} \ding{121} \ding{121} \ding{121} \ding{121} \ding{121} \ding{121} \ding{121} \ding{121} \ding{121} \ding{121} \ding{121} \ding{121} \ding{121} \ding{121} \ding{121} \ding{121} \ding{121}}}&\cellcolor{gray}\textcolor{gray}{-}0.66\textcolor{gray}{..}\textcolor{blue2}{  \ding{108}}&\multicolumn{1}{c}{\textcolor{g1}{\ding{121} \ding{121} \ding{121} \ding{121} \ding{121} \ding{121} \ding{121} \ding{121} \ding{121} \ding{121} \ding{121}}}&\multicolumn{1}{c|}{\ding{121}}&\textcolor{white}{-}0.22&\multicolumn{1}{c}{\ding{121}}&\multicolumn{1}{c|}{\ding{121}}\\

\cellcolor{g1}{\bf TL-TS}&\textcolor{white}{-}0.32&\multicolumn{1}{c}{\ding{121}}&\multicolumn{1}{c|}{\ding{121}}&\textcolor{white}{-}0.31&\multicolumn{2}{c|}{\textcolor{g1}{\ding{121} \ding{121} \ding{121} \ding{121} \ding{121} \ding{121} \ding{121} \ding{121} \ding{121} \ding{121} \ding{121} \ding{121} \ding{121} \ding{121} \ding{121} \ding{121} \ding{121} \ding{121} \ding{121} \ding{121} \ding{121} \ding{121} \ding{121}}}&\textcolor{white}{-}0.33&\multicolumn{2}{c|}{\textcolor{g1}{\ding{121} \ding{121} \ding{121} \ding{121} \ding{121} \ding{121} \ding{121} \ding{121} \ding{121} \ding{121} \ding{121} \ding{121} \ding{121} \ding{121} \ding{121} \ding{121} \ding{121} \ding{121} \ding{121} \ding{121} \ding{121} \ding{121} \ding{121}}}&-0.41&\multicolumn{1}{c}{\textcolor{g1}{\ding{121} \ding{121} \ding{121} \ding{121} \ding{121} \ding{121} \ding{121} \ding{121} \ding{121} \ding{121} \ding{121} }}&\multicolumn{1}{c|}{\ding{121}}&\cellcolor{gray}\textcolor{gray}{-}0.63\textcolor{gray}{..}\textcolor{blue2}{  \ding{108}}&\multicolumn{1}{c}{\ding{121}}&\multicolumn{1}{c|}{\ding{121}}\\\hline\hline

{\bf TD-IT}&\cellcolor{gray}\textcolor{gray}{-}0.54\textcolor{gray}{..}\textcolor{blue1}{  \ding{108}}&\multicolumn{1}{c}{\ding{121}}&\multicolumn{1}{c|}{\ding{121}}&\cellcolor{gray}\textcolor{gray}{-}0.82\textcolor{gray}{..}\textcolor{blue3}{  \ding{108}}&\multicolumn{1}{c}{\cellcolor{gray}3e-4 \(\langle 00, 11 \rangle\)}&&\cellcolor{gray}\textcolor{gray}{-}0.74\textcolor{gray}{..}\textcolor{blue2}{  \ding{108}}&&&\cellcolor{gray}\textcolor{gray}{-}0.73\textcolor{gray}{..}\textcolor{blue2}{  \ding{108}}&&\multicolumn{1}{c|}{\ding{121}}&\cellcolor{gray}\textcolor{gray}{-}0.89\textcolor{gray}{..}\textcolor{blue3}{  \ding{108}}&\multicolumn{1}{c}{\ding{121}}&\multicolumn{1}{c|}{\ding{121}}\\
{\bf TD-PC}&\textcolor{white}{-}0.05&\multicolumn{1}{c}{\ding{121}}&\multicolumn{1}{c|}{\ding{121}}&\cellcolor{gray}-0.69\textcolor{gray}{..}\textcolor{red1}{  \ding{108}}&&&-0.10&\cellcolor{gray}0.01 \(\langle 11, 00 \rangle\)&&\cellcolor{gray}-0.60\textcolor{gray}{..}\textcolor{red1}{  \ding{108}}&&\multicolumn{1}{c|}{\ding{121}}&\cellcolor{gray}\textcolor{gray}{-}0.61\textcolor{gray}{..}\textcolor{blue1}{  \ding{108}}&\multicolumn{1}{c}{\ding{121}}&\multicolumn{1}{c|}{\ding{121}}\\

{\bf TD-DT}&-0.01&\multicolumn{1}{c}{\ding{121}}&\multicolumn{1}{c|}{\ding{121}}&\textcolor{white}{-}0.36&\multicolumn{1}{c}{\cellcolor{gray}0.02 \(\langle10, 01\rangle\)}&\multicolumn{1}{c|}{\cellcolor{gray}3e-4 \(\langle 00, 11 \rangle\)}&\textcolor{white}{-}0.14&\cellcolor{gray}0.01 (10, 01)&\multicolumn{1}{c|}{\cellcolor{gray}0.02 \(\langle 10, 01 \rangle\)}&\textcolor{white}{-}0.28&\multicolumn{1}{c}{\cellcolor{gray}0.01 \(\langle10, 01\rangle\)}&\multicolumn{1}{c|}{\ding{121}}&-0.24&\multicolumn{1}{c}{\ding{121}}&\multicolumn{1}{c|}{\ding{121}}\\

{\bf TD-TS}&\textcolor{white}{-}0.27&\multicolumn{1}{c}{\ding{121}}&\multicolumn{1}{c|}{\ding{121}}&-0.20&&&\cellcolor{gray}-0.55\textcolor{gray}{..}\textcolor{red1}{  \ding{108}}&\cellcolor{gray}0.02 \(\langle11, 00\rangle\)&&\cellcolor{gray}-0.64\textcolor{gray}{..}\textcolor{red1}{  \ding{108}}&&\multicolumn{1}{c|}{\ding{121}}&\cellcolor{gray}-0.91\textcolor{gray}{..}\textcolor{red3}{  \ding{108}}&\multicolumn{1}{c}{\ding{121}}&\multicolumn{1}{c|}{\ding{121}}\\\hline\hline

\cellcolor{g1}{\bf IT-PC}&-0.21&\multicolumn{1}{c}{\ding{121}}&\multicolumn{1}{c|}{\ding{121}}&\cellcolor{gray}-0.76\textcolor{gray}{..}\textcolor{red2}{  \ding{108}}&\multicolumn{2}{c|}{\textcolor{g1}{\ding{121} \ding{121} \ding{121} \ding{121} \ding{121} \ding{121} \ding{121} \ding{121} \ding{121} \ding{121} \ding{121} \ding{121} \ding{121} \ding{121} \ding{121} \ding{121} \ding{121} \ding{121} \ding{121} \ding{121} \ding{121} \ding{121} \ding{121}}}&\cellcolor{gray}-0.53\textcolor{gray}{..}\textcolor{red1}{  \ding{108}}&\multicolumn{2}{c|}{\textcolor{g1}{\ding{121} \ding{121} \ding{121} \ding{121} \ding{121} \ding{121} \ding{121} \ding{121} \ding{121} \ding{121} \ding{121} \ding{121} \ding{121} \ding{121} \ding{121} \ding{121} \ding{121} \ding{121} \ding{121} \ding{121} \ding{121} \ding{121} \ding{121}}}&\cellcolor{gray}-0.94\textcolor{gray}{..}\textcolor{red3}{  \ding{108}}&\multicolumn{1}{c}{\textcolor{g1}{\ding{121} \ding{121} \ding{121} \ding{121} \ding{121} \ding{121} \ding{121} \ding{121} \ding{121} \ding{121} \ding{121}}}&\multicolumn{1}{c|}{\ding{121}}&\textcolor{gray}{-}0.61&\multicolumn{1}{c}{\ding{121}}&\multicolumn{1}{c|}{\ding{121}}\\

{\bf IT-DT}&\textcolor{white}{-}0.10&\multicolumn{1}{c}{\ding{121}}&\multicolumn{1}{c|}{\ding{121}}&\cellcolor{gray}\textcolor{gray}{-}0.63\textcolor{gray}{..}\textcolor{blue2}{  \ding{108}}&&\multicolumn{1}{c|}{\cellcolor{gray}0.02 \(\langle 00, 11 \rangle\)}&\cellcolor{gray}\textcolor{gray}{-}0.52\textcolor{gray}{..}\textcolor{blue1}{  \ding{108}}&&&\cellcolor{gray}\textcolor{gray}{-}0.83\textcolor{gray}{..}\textcolor{blue3}{  \ding{108}}&&\multicolumn{1}{c|}{\ding{121}}&-0.15&\multicolumn{1}{c}{\ding{121}}&\multicolumn{1}{c|}{\ding{121}}\\

\cellcolor{g1}{\bf IT-TS}&-0.31&\multicolumn{1}{c}{\ding{121}}&\multicolumn{1}{c|}{\ding{121}}&-0.41&\multicolumn{2}{c|}{\textcolor{g1}{\ding{121} \ding{121} \ding{121} \ding{121} \ding{121} \ding{121} \ding{121} \ding{121} \ding{121} \ding{121} \ding{121} \ding{121} \ding{121} \ding{121} \ding{121} \ding{121} \ding{121} \ding{121} \ding{121} \ding{121} \ding{121} \ding{121} \ding{121}}}&-0.36&\multicolumn{2}{c|}{\textcolor{g1}{\ding{121} \ding{121} \ding{121} \ding{121} \ding{121} \ding{121} \ding{121} \ding{121} \ding{121} \ding{121} \ding{121} \ding{121} \ding{121} \ding{121} \ding{121} \ding{121} \ding{121} \ding{121} \ding{121} \ding{121} \ding{121} \ding{121} \ding{121}}}&\cellcolor{gray}-0.86\textcolor{gray}{..}\textcolor{red3}{  \ding{108}}&\multicolumn{1}{c}{\textcolor{g1}{\ding{121} \ding{121} \ding{121} \ding{121} \ding{121} \ding{121} \ding{121} \ding{121} \ding{121} \ding{121} \ding{121}}}&\multicolumn{1}{c|}{\ding{121}}&\cellcolor{gray}-0.73\textcolor{gray}{..}\textcolor{red2}{  \ding{108}}&\multicolumn{1}{c}{\ding{121}}&\multicolumn{1}{c|}{\ding{121}}\\\hline\hline

\cellcolor{g1}{\bf PC-DT}&\cellcolor{gray}\textcolor{gray}{-}0.50\textcolor{gray}{..}\textcolor{blue1}{  \ding{108}}&\multicolumn{1}{c}{\ding{121}}&\multicolumn{1}{c|}{\ding{121}}&\cellcolor{gray}-0.62\textcolor{gray}{..}\textcolor{red1}{  \ding{108}}&\multicolumn{2}{c|}{\textcolor{g1}{\ding{121} \ding{121} \ding{121} \ding{121} \ding{121} \ding{121} \ding{121} \ding{121} \ding{121} \ding{121} \ding{121} \ding{121} \ding{121} \ding{121} \ding{121} \ding{121} \ding{121} \ding{121} \ding{121} \ding{121} \ding{121} \ding{121} \ding{121}}}&-0.16&\multicolumn{2}{c|}{\textcolor{g1}{\ding{121} \ding{121} \ding{121} \ding{121} \ding{121} \ding{121} \ding{121} \ding{121} \ding{121} \ding{121} \ding{121} \ding{121} \ding{121} \ding{121} \ding{121} \ding{121} \ding{121} \ding{121} \ding{121} \ding{121} \ding{121} \ding{121} \ding{121}}}&\cellcolor{gray}-0.78\textcolor{gray}{..}\textcolor{red2}{  \ding{108}}&\multicolumn{1}{c}{\textcolor{g1}{\ding{121} \ding{121} \ding{121} \ding{121} \ding{121} \ding{121} \ding{121} \ding{121} \ding{121} \ding{121} \ding{121}}}&\multicolumn{1}{c|}{\ding{121}}&\cellcolor{gray}-0.63\textcolor{gray}{..}\textcolor{red1}{  \ding{108}}&\multicolumn{1}{c}{\ding{121}}&\multicolumn{1}{c|}{\ding{121}}\\

{\bf PC-TS}&\textcolor{white}{-}0.35&\multicolumn{1}{c}{\ding{121}}&\multicolumn{1}{c|}{\ding{121}}&\textcolor{white}{-}0.10&&&-0.47&&\multicolumn{1}{c|}{\cellcolor{gray}0.01 \(\langle 10, 01 \rangle\)}&\cellcolor{gray}\textcolor{gray}{-}0.85\textcolor{gray}{..}\textcolor{blue3}{  \ding{108}}&&\multicolumn{1}{c|}{\ding{121}}&\cellcolor{gray}-0.59\textcolor{gray}{..}\textcolor{red1}{  \ding{108}}&\multicolumn{1}{c}{\ding{121}}&\multicolumn{1}{c|}{\ding{121}}\\\hline\hline

\cellcolor{g1}{\bf DT-TS}&\textcolor{white}{-}0.36&\multicolumn{1}{c}{\ding{121}}&\multicolumn{1}{c|}{\ding{121}}&-0.41&\multicolumn{2}{c|}{\textcolor{g1}{\ding{121} \ding{121} \ding{121} \ding{121} \ding{121} \ding{121} \ding{121} \ding{121} \ding{121} \ding{121} \ding{121} \ding{121} \ding{121} \ding{121} \ding{121} \ding{121} \ding{121} \ding{121} \ding{121} \ding{121} \ding{121} \ding{121} \ding{121}}}&\cellcolor{gray}-0.54\textcolor{gray}{..}\textcolor{red1}{  \ding{108}}&\multicolumn{2}{c|}{\textcolor{g1}{\ding{121} \ding{121} \ding{121} \ding{121} \ding{121} \ding{121} \ding{121} \ding{121} \ding{121} \ding{121} \ding{121} \ding{121} \ding{121} \ding{121} \ding{121} \ding{121} \ding{121} \ding{121} \ding{121} \ding{121} \ding{121} \ding{121} \ding{121}}}&\cellcolor{gray}-0.68\textcolor{gray}{..}\textcolor{red2}{  \ding{108}}&\multicolumn{1}{c}{\textcolor{g1}{\ding{121} \ding{121} \ding{121} \ding{121} \ding{121} \ding{121} \ding{121} \ding{121} \ding{121} \ding{121} \ding{121} }}&\multicolumn{1}{c|}{\ding{121}}&\cellcolor{gray}\textcolor{gray}{-}0.52\textcolor{gray}{..}\textcolor{blue1}{  \ding{108}}&\multicolumn{1}{c}{\ding{121}}&\multicolumn{1}{c|}{\ding{121}}\\\hline

\multicolumn{8}{l}{{\textcolor{blue1}{ \scriptsize \ding{108}} : \(0.50\leq \rho < 0.65\)}, \textcolor{white} {..........}{\textcolor{blue2}{ \scriptsize \ding{108}} : \(0.65 \leq \rho <0.80\)}, \textcolor{white} {...........}{\textcolor{blue3}{ \scriptsize \ding{108}} : \(\rho \geq 0.80\)},}&\multicolumn{8}{l}{\textcolor{g1}{\ding{121} \ding{121} \ding{121} \ding{121} \ding{121}}: {\it No interaction}}\\
\multicolumn{8}{l}{{\textcolor{red1}{ \scriptsize \ding{108}} : \(-0.65 < \rho \leq -0.50\)}, {\textcolor{white} {.....}\textcolor{red2}{ \scriptsize \ding{108}} : \(-0.80 < \rho \leq -0.65\)}, {\textcolor{white} {.....}\textcolor{red3}{ \scriptsize \ding{108}} : \(\rho \leq -0.80\)}}\\
\end{tabular}
\vspace{-3mm}
\end{table*}
%----------------------------------------------------------
\vspace{-2mm}
\subsection{RQ3- Two-way Impact}
We consider cross-factor correlations to assess the relationship strength among \(iv_{1-8}\). 
Since all of the independent variables of our repository analysis are recorded in a binary format, the Phi coefficient test is used to determine the degree and the strength of association between these variables. 
We then analyze the two-way interaction of these factors on the disruptiveness of interruptions in software development tasks (see Figure~\ref{fig:Interaction}). 
The gray-highlighted cells in Table~\ref{tab:RQ3} show the correlation and the interaction between each pair of factors, and the colored circles denote the strength of these correlations. 

{\bf Finding \(_{\bf 3-1}\):} The Phi correlation tests show that for all of the task types studied, there is a significant positive correlation ({\(\phi >\)}0.50, df=1,  \(\chi^2    >\)10.8,  \(p<\)0.001) between \emph{type difference} and \emph{interruption type} factors. 
This implies that self-initiated task switchings are mainly associated with a change in the task type. Moreover, in all task types except testing, context switching and experience level variables are negatively correlated with the task level (CS: \(\phi\leq -0.64\), df=1,  \(\chi^2    >\)10.8,  \(p<0.001\); EL: \(\phi\leq -0.80\), df=1,  \(\chi^2    >\)10.8, \(p<0.001\)), indicating that for more experienced developers task or context switching are usually high-level tasks. 
 
{\bf Finding \(_{\bf 3-2}\):} Regarding the interruption timing, there is a significant positive correlation between interruption type and daytime variables for programming, testing, and UI design tasks (\(\phi\geq0.52\),  \(p<0.001\)). This implies self-initiated interruptions usually happen in the morning. In addition, self-interruptions are associated with interruptions characterized by a priority change (\(\phi\leq -0.53\)).
 
{\bf Finding \(_{\bf 3-3}\):} Table \ref{tab:RQ3} (row TD-DT) shows the interaction between \emph{type change} and \emph{daytime} variables significantly (i.e. SRH tests) impacts both disruptive factors of programming and testing tasks and suspension period of UI task interruptions. 
For all these three task types, the \(\langle 10, 01 \rangle\) (i.e. \(\langle\) different type/morning, same type/afternoon\(\rangle\)) combination negatively impacts the suspension period and for programming and testing tasks the \(\langle 00, 11 \rangle\) (i.e. \(\langle\)different type/afternoon, same type/morning\(\rangle\)) negatively impacts the nested interruption parameters. 

{\bf Finding \(_{\bf 3-4}\):} The interaction between task level and type difference variables significantly impacts the disruptiveness of programming and UI interruptions. This interaction is more disruptive when the task switching is characterized as \(\langle\)main-task/different type, sub-task/same type\(\rangle\)).
  
{\bf Finding \(_{\bf 3-5}\):} While \emph{experience level} alone does not make any significant difference on the disruptiveness of programming tasks (Tables \ref{tab:RQ2}, p\(>\)0.05), when it interacts with \emph{type difference, interruption type}, or \emph{priority change} these variables significantly impact interruptions to this task type. 
For example, \(\langle01, 01\rangle = \langle\)less exp/self-int, more exp/external-int\(\rangle\) negatively impact programming task interruptions. Likewise, \emph{context switching} alone does not impact interruptions in testing tasks, but its interaction with \emph{type difference, interrutption type}, or \emph{priority change} does. 
    
{\bf Discussion \(_{\bf 3-1}\):} We applied Spearman's rank test on survey responses to questions about the disruptiveness of various interruptions characteristics (see Figure \ref{fig:Likert1}). 
The results reveal that there is a weak correlation between context and type switching variables (i.e. CS/TD: rho=0.2, \(p\)=0.04). 
This shows that respondents who rated context switching as a disruptive interruption factor, did so for the type switching factor: \emph{``Changing a task type is disruptive if it made me change environment e.g. Launch different servers''}. 
Spearman's rank tests also show a weak correlation between type difference (TD) and each of interruption type (IT) and task stage (TS) factors (TD/TS: rho= 0.19, \(p\)=0.04; TD/IT: rho=0.22, \(p\)=0.02), as in: \emph{``the disruptiveness of type switching depends on if I reached a good stopping point before the switch or not [...]''}. 
We found there is a correlation between participants' rating to the disruptiveness of context switching (CS) and interruption type (IT) factors (CS/IT: rho=0.37, \(p\)=1e-7). 
Similar to the results of our retrospective interaction analysis, respondents who rated context switching as a disruptive factor found external interruptions more disruptive than self-interruptions: \emph{``typically the interruptions that come from others are longer reaching - often it means that my skills are needed elsewhere, and so I need to switch tasks or projects for a more extended period, which adds more items to my cognitive stack''}.

{\bf Discussion \(_{\bf 3-2}\):} We propose a set of correlation and interaction patterns that can be used to interpret developers' task switching behaviour and to investigate the cross-factor impact of task switching characteristics. 
We present these patterns as: 
    % [noitemsep,topsep=0pt]
    \vspace{-1mm}
  {  \begin{description}  [labelsep=0.1em, labelwidth=0.35in, labelindent=0cm, align=left]
  \item   [\emph{Correlation Patterns:}] \( \langle T,(\imath\nu_i, \imath\nu_j),\) {\small \ding{108} }\(\rangle\), where \(i\) and \(j \in \{1, 2, ..., 8\}\) and denote two distinct interruption characteristics and the color of {\small \ding{108} } presents the direction and the strength of the association between these characteristics. For instance, \(\langle \text{\it Programming}, (CS, TL),\) {\small \textcolor{red3}{\ding{108}} }\(\rangle\) indicates there is a strong negative association between the context switching and task level variables in programming tasks' interruptions.\\[-1em]
  \item [Interaction Patterns:] \(\big< (\imath\nu_i, \imath\nu_j), (\alpha \beta, \bar{\alpha}\bar{\beta} ), {\Delta/|w|}_T\big>\), which implies the interaction between two distinct interruption characteristics \(\imath\nu_i\) and \(\imath\nu_j\) with the values of \(\alpha\) and  \(\beta\) (i.e. \(\alpha, \beta \in \{0,1\}\)) negatively impacts \(\Delta\) and/or \(|w|\) of task \(T\)'s interruptions. For instance, \(\big< (TD, PC), (11,00), {\Delta}_{Test}\big>\) indicates that (diff type/same priority, same type/diff priority) significantly impact interruptions of testing tasks and negatively impact their suspension period.
    \end{description}}
     
These patterns along with the detailed information presented in Table \ref{tab:RQ3}, can be used to guide decision-making and forecasting the consequences of task switching decisions.
     
     {\bf Practitioner's corner \(_{\bf 3}\):} While there are various combinations of factors which can impact the disruptiveness of interruptions in a negative way, the results of this section do not exactly prove that interruptions are always disruptive. There are circumstances where task switching or interruptions can boost developers' productivity, as stated by one of our survey respondents: \emph{``Learning takes time. Sometimes I learn basics for a task then I leave it for the next day which makes me mentally prepared for the task. Or, if a team member asks me a question about a portion of a feature which they are working on, that often gives me clarity about what I am working one''}. We propose that task switching is a skill and not an obstacle to work. Designing the development processes in a way to be resilient to interruptions can mitigate the risk of unplanned and disruptive interruptions. For instance, having frequent, small commits help a team keep the amount of work that they have not yet submitted always very small.  Mapping each commit to one discrete change to the source code (e.g. refactoring, a failing test, or a TDD cycle) and encoding all of developers' knowledge about the code into the code itself (e.g. by extracting methods and renaming methods and variables to reflect their meaning) help reduce the cognitive cost of unavoidable task switching and interruptions occur to programming tasks.     %
\section{Threats to Validity}
Although our longitudinal study used data collected from a  single company we argue that our findings generalize.
We tried to mitigate this risk by implementing our repository study on a fairly large dataset including various projects from different business domains and employees from different levels of experience. 
Our data collection and preparation pose another threat to the validity of our results because identifying the interruption type (i.e. self and external) and temporal stage of tasks (i.e. early and late) is not straightforward. 
The pilot studies we conducted before our main data collection phase helped address this risk. 
Additionally, the retrospective dataset associated with each employee was reviewed by at least two hired RA's and the first author of the paper. 
To evaluate the reliability of our decisions for independent variables that have been recorded manually, we used the Cohen's Kappa statistic, which calculates the degree of agreement between two evaluators. 
The calculated Kappa value was 0.87, which shows significant agreement according to  Landis and Koch \cite{Kappa}. 
In regard to survey results, we pilot tested the survey questions with three software developers to mitigate the risk of misunderstanding questions.
However, the questions still require participant interpretation. 
We mitigate this risk by adding a comment space for each question and asked respondents to clarify their response or discuss other aspects of the question if they desired. 
The survey population could be biased towards a specific population so the generalizability of our survey results may have intrinsic limits. 
We mitigate this by distributing our survey to a large number of potential respondents with different levels of software development experience and from various countries (e.g. Germany, Netherland, Sweden, Hungary, USA, New Zealand, and Canada).

%(+)(+)(+)(+)(+)(+)(+)(+)(+)(+)(+)(+)(+)(+)(+)(+)(+)(+)(+)(+)(+)(+)(+)(+)(+)(+)(+)

%(+)(+)(+)(+)(+)(+)(+)(+)(+)(+)(+)(+)(+)(+)(+)(+)(+)(+)(+)(+)(+)(+)(+)(+)(+)(+)(+)

%=================================================================================
\vspace{-2mm}
\section{Conclusion and Implications}
Interruption, as a form of task switching or sequential multitasking, is an inherent part of software development tasks. Not all of the interruptions should be counted as waste because in some specific cases task switching is unavoidable and can actually increase developers' productivity. 
Using a mixed-methods study including a retrospective analysis and a survey, we studied the disruptive impact of various interruption characteristics on development tasks interruptions. We found that the problem-solving nature of programming and testing tasks make them more vulnerable to interruptions compared to architecture and UI design tasks. Interestingly, we found self-interruptions negatively impact the disruptiveness of interruptions in all types of development tasks. However,  the survey responses reveal that developers seem to believe external-interruptions are more vulnerable than self-interruptions. 
We also provided a set of recommendations (see {\em practitioners' corners}) for project managers and practitioners which can be used as a mean to guide decision-making and forecasting the consequences of task switching decisions in software development teams.

We suggest that research in multitasking and task interruptions in the area of software engineering focus on measuring and characterizing the cost of task switching and interruptions. As the differences between our repository analysis and survey data reveal and as supported by recent practical studies (see  \cite{Sky, Pro1, Meyer2017}), the disruptiveness of task switching is most likely to be affected by the context in which the switching occurs. 
As one of the respondents said: \emph{``[...] If someone is working on the same project as I am and we can exchange ideas, that can be a productive task switching. It's also productive for more fire-drill type situations, like fast bug triage.''}